\documentclass[twocolumn,trackchanges,floatfix]{aastex631}

\usepackage{amsmath}

\received{2024 January 05}
\revised{2024 February 24}
\accepted{2024 February 26}

\graphicspath{{./}{figures/}}

\begin{document}

\title{Exploring NGC 2345: A Comprehensive Study of a Young Open Cluster through Photometric and Kinematic Analysis}

\author{Kuldeep Belwal}
\affiliation{Indian Centre for Space Physics,466 Barakhola, Singabari road, Netai Nagar,
Kolkata, 700099, India}

\author[0000-0002-8988-8434]{D. Bisht}
\affiliation{Indian Centre for Space Physics,466 Barakhola, Singabari road, Netai Nagar,
Kolkata, 700099, India}


\author{Mohit Singh Bisht}
\affiliation{Indian Centre for Space Physics,466 Barakhola, Singabari road, Netai Nagar,
Kolkata, 700099, India}

\author[0000-0002-6373-770X]{Geeta Rangwal}
\affiliation{South-Western Institute for Astronomy Research, Yunnan University, Kunming 650500, People’s Republic of China}

\author{Ashish Raj}
\affiliation{Indian Centre for Space Physics,466 Barakhola, Singabari road, Netai Nagar,
Kolkata, 700099, India}

\author[ 0000-0002-4729-9316]{Arvind K. Dattatrey}
\affiliation{Aryabhatta Research Institute of Observational Sciences, Manora Peak, Nainital 263129, India}

\author[0000-0003-1342-9626]{R. K. S. Yadav}
\affiliation{Aryabhatta Research Institute of Observational Sciences, Manora Peak, Nainital 263129, India}

\author{B. C. Bhatt}
\affiliation{Indian Institute of Astrophysics, 560034 Bangalore, India}

\email{devendrabisht297@gmail.com}
\email{kuldeepbelwal1997@gmail.com}
\email{rkant@aries.res.in}
\email{physics.arvind97@gmail.com}
\begin{abstract}

We conducted a photometric and kinematic analysis of the young open cluster NGC 2345 using CCD \emph{UBV} data from 2-m Himalayan Chandra Telescope (HCT), \emph{Gaia} Data Release 3 (DR3), 2MASS, and the APASS datasets. We found 1732 most probable cluster members with membership probability higher than 70$\%$. The fundamental and structural parameters of the cluster are determined based on the cluster members. The mean proper motion of the cluster is estimated to be $\mu_{\alpha}cos\delta$ = ${-1.34}\pm0.20$ and $\mu_{\delta}$= $1.35\pm 0.21$ mas $yr^{-1}$. Based on the radial density profile, the estimated radius is $\sim$ 12.8 arcmin (10.37 pc). Using color-color and color-magnitude diagrams, we estimate the reddening, age, and distance to be $0.63\pm0.04$ mag, 63 $\pm$ 8 Myr, and 2.78 $\pm$ 0.78 kpc, respectively. The mass function slope for main-sequence stars is determined as $1.2\pm 0.1$. The mass function slope in the core, halo, and overall region indicates a possible hint of mass segregation. The cluster's dynamical relaxation time is 177.6 Myr, meaning ongoing mass segregation, with complete equilibrium expected in 100-110 Myr. Apex coordinates are determined as  $-40^{\circ}.89 \pm 0.12, -44^{\circ}.99 \pm 0.15$. The cluster's orbit in the Galaxy suggests early dissociation in field stars due to its close proximity to the Galactic disk.

\end{abstract}


\section{Introduction} 

\label{sec:Introduction}
A star cluster comprises a collection of stars, providing a natural laboratory-like setting to develop theoretical models and apply them to gain insights into stellar evolution. Open Clusters (OCs) are gatherings of stars originating from the same molecular cloud, characterized by the same age, chemical composition, and distance; however, stars' luminosity and masses are different \citep{yontan2023investigation}. Typically composed of a few tens to several thousand stars, they loosely aggregate and remain gravitationally bound to each other. In our Galaxy, most, if not all, stars are born in clusters \citep{portegies2010young}. The study of star clusters is crucial to understand star formation and stellar evolution. OCs are relatively young systems, witnessing recent star formation events, and are predominantly located in the spiral arms of the Milky Way. Young open clusters are typically situated within the densely populated environment of the Galactic disk. Therefore, it is essential to differentiate cluster members from field stars to estimate their physical parameters accurately \cite{carraro2008old}; \cite{dias2018astrometric}. Numerous recent studies have conducted membership analyses of stars in the vicinity of open clusters and have explored various cluster properties (\cite{bisht2019mass},\cite{bisht2020comprehensive};  \cite{castro2018new},\cite{castro2019hunting}; \cite{cantat2018gaia}; \cite{liu2019catalog}). The release of \emph{Gaia DR2} data in 2018 marked a revolution in astronomy, providing unprecedented levels of accuracy in astrometry data (\cite{cantat2019gaia}; \cite{monteiro2019distances}). Membership helps  to study the distribution of luminosity and mass during star formation, which is known as the initial mass function (IMF) (\cite{maurya2020photometric}). Whether the IMF is universal in time and space or depends on various star-forming conditions remains a subject of debate (\cite{bastian2010universal};\cite{dib2018emergence}). \\ 
The Young open clusters are essential tools to study the effect of mass segregation, where massive stars are concentrated towards the center, and lower mass stars spread in the outer region. Whether mass segregation in star clusters is primarily due to the dynamical evolution of the cluster or the star formation process itself is a subject of ongoing research and debate in astrophysics \citep{dib2018structure}.\\
\vspace{-20pt} %
\begin{figure}[th]
    \includegraphics[width=9.5cm,height=7.2cm]{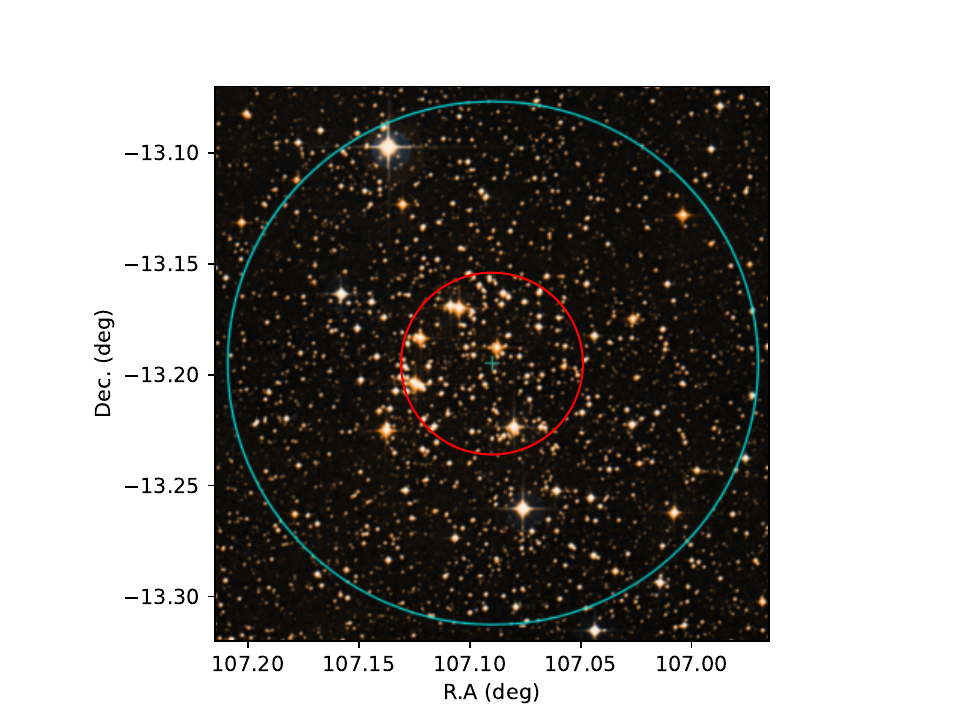}
    \caption{Identification chart of NGC 2345 of size 15$\times$15 arcmin$^2$ taken from the DSS and oriented in the northeast direction (North is up and East is left). The inner circle represents the core radius (3.9 arcmin), while the outer ring denotes the cluster radius (12.8 arcmin).}
    \label{fig:Identification chart}
\end{figure}

The young open cluster NGC 2345 (C0706$-$130) is located in the constellation of Canis Major [$\alpha_{2000}=07^h 08^m 18^s, \delta_{ 2000}=-13^{\circ}11^{'}36^{"}$] and corresponding Galactic coordinate [$l=226^{\circ}.58, b=-2^{\circ}.31$] given in the WEBDA open cluster dataset{\footnote{\url{https://webda.physics.muni.cz/}}}. The cluster identification chart is taken from the Digitized Sky Survey (DSS)\footnote{\url{https://simbad.u-strasbg.fr/simbad/}} and shown in Fig. \ref{fig:Identification chart}. The estimated distance, reddening $(E(B-V))$, and age of this cluster fall within the ranges of $2.2 - 3.0$ kpc, 0.59-0.68 mag, and 55-79 Myr (\cite{kharchenko2005astrophysical}, \cite{kharchenko2013global}; \cite{carraro2014thickening}; \cite{cantat2018gaia};  \cite{alonso2019comprehensive},\cite{singh2022polarization}). \cite{kharchenko2005astrophysical} derived the core and cluster radius as $4'.2$ and $7'.2$ respectively, whereas \cite{alonso2019comprehensive} estimated core radius 3.44 $\pm$ 0.08 arcmin, tidal radius $18.7  \pm  1.2 $ arcmin and metallicity $[Fe/H]= -0.28 $ for the cluster. In various past studies, a non-radial distribution of dust associated with the cluster was observed, with variable reddening $(E(B - V))$ ranging from 0.4 to 1.2 mag (\cite{moffat1974ngc}; \cite{carraro2014thickening}; \cite{alonso2019comprehensive}). We used the radial velocity $58.41\pm 0.15$ km/s provided by \cite{dias2021updated} and \cite{carrera2022one} in the further analysis.\\

Previous studies of cluster NGC 2345 have shown a range in the fundamental parameters. However, these parameters were not determined using the cluster members. Thanks to the  \emph{Gaia} DR3 survey, we now have access to the much more precise proper motion of the stars in this cluster. These proper motions have been used to calculate the membership probability, which helps us identify the members of the clusters. For the first time, we analyzed the cluster members to determine the fundamental parameters of cluster NGC 2345. This cluster presents a valuable opportunity to study stellar evolution as it contains blue and red supergiants, has low metallicity, and has a high fraction of Be stars \citep{alonso2019comprehensive}. However, further detailed analysis is needed due to the peculiarity of this cluster and the lack of studies on its members. 

Estimating the orbital parameters of the young open cluster NGC 2345 is pivotal for unraveling its historical and future trajectory within the Milky Way. We gain insights into the underlying galactic dynamics and gravitational forces influencing this object by determining the cluster's orbit shape, orientation, and velocity. This information contributes to understand the cluster's formation and evolutionary history and enlightens its interactions with other structures in the Milky Way and the conditions prevalent during its birth. The derived orbital parameters serve as a crucial benchmark for meticulous comparisons with theoretical models, refining our understanding of galactic dynamics and extending our more comprehensive knowledge of the evolutionary processes governing young open clusters in our galaxy.

This analysis presents a photometric study for the open cluster NGC 2345. We investigated the spatial structure, fundamental parameters, extent, reddening, and age of the cluster, aiming to gain insights into its dynamical evolution. This analysis is based on kinematic and CCD photometric data extending to approximately $V = 19$ mag. We estimated membership probabilities for stars within the NGC 2345 cluster down to $G \sim 20$ mag. We used high-probability members to analyze the mass function and mass segregation. The paper is structured as follows: Section~\ref{sec:Observations and data reduction} presents the observations and data analysis. Section~\ref{sec:Membership probability} discusses the stellar membership within the clusters. In Section~\ref{sec:Structural Parameters of the Cluster}, we derive the structural parameters of the clusters, followed by the estimation of physical parameters in Section~\ref{sec:DISTANCE AND AGE OF THE CLUSTER}. Section~\ref{sec:Dynamical Study} explores the dynamical evolution of the clusters. Finally, our work is summarized in Section~\ref{sec:Summary and conclusion}. \\

\section{Observations and Data Reduction}
\label{sec:Observations and data reduction}
The photometric observations for this cluster were conducted using the 2.0-m Himalayan Chandra Telescope (HCT) at Hanle, operated by the Indian Institute of Astrophysics, Bangalore, India. The Hanle Faint Object Spectrograph Camera (HFOSC), which is an imager cum spectrograph, was used for photometric observation of NGC 2345 in \emph{UBV} Bessel filters. The detector is a 2K $\times$ 4K CCD, where the central 2K $\times$ 2K pixels were used for imaging. The pixel size is 15  $\mu$m with an image scale of 0.297 arcsec /pixel. The field of view for imaging is approximately 10 $\times$ 10 arcmin$^{2}$, readout noise 4.8 $e^{-}$ and gain 1.22 e$^{-}$/ADU. The observations have a typical mean S/N ratio of $>$ 1000. The observing log is given in Table \ref{tab: photometric observations}. We observed many bias and twilight flat-field frames in the \emph{UBV} filters and multiple short and long exposure frames of the target and standard fields at night. Standard field PG 1525 of Landolt (1992) was observed to calibrate the photometry of cluster stars. To perform the initial processing of the raw data, we utilized the IRAF \footnote[2]{The Image Reduction and Analysis Facility (IRAF) is distributed by the National Optical Astronomy Observatories (NOAO).} data reduction packages, which include bias subtraction, flat fielding, and cosmic ray removal. The instrumental magnitudes were estimated through point spread function (PSF) fitting using DAOPHOT II (\cite{stetson1987daophot}, \cite{worrall1992astronomical}) package. 

\begin{table}
\caption{Log of observations, with dates, exposure times, and air masses for each passband. Data were taken on 02/03 February 2004.}
\begin{center}
\begin{tabular}{|c|c|c|c|}
\hline
 Cluster/ &Filters & Exp.(s)× no. of & Airmass \\
 Standard Field  &  &  frames        & \\
\hline
 NGC 2345 & {\textit{V}}   &  50 x 2     & 1.61 - 1.66  \\
          &     &  30 x 3                &      \\

   & {\textit{B}} & 100 x 3  & 1.56 - 1.58 \\

  & {\textit{U}} & 180 x 3   & 1.51 - 1.53 \\
\hline
 PG 1525 & {\textit{V}}   &  80 x 5    & 1.30 - 1.52 \\

   &  {\textit{B}} & 150 x 6   & 1.32 - 1.59 \\

  & {\textit{U}} &  300 x 6    & 1.31 - 1.61 \\
  
\hline
\end{tabular}
\end{center}
\label{tab: photometric observations}
\end{table} 
\subsection{Photometric calibration} 

We also observed the standard field PG 1525 \citep{landolt1992ubvri} during the same observing night for photometric calibration. The five standard stars (PG 1525-071, 071A, 071B, 071C, 071D ) used in the calibration have brightness and  color range 9$\leq V \leq$19 and  0.3$\leq$ $B-V \leq$ 2.0 respectively. For the extinction coefficients, we assumed the typical values for the Hanle observing site \citep{stalin2008night}. For translating the instrumental magnitude to the standard magnitude, the calibration equations using the least squares linear regression are as follows:

\begin{equation}
    v = V + Z_V +C_V(B-V) + k_V X
\end{equation}
\begin{equation}
    b = B + Z_B +C_B(B-V) + k_B X
\end{equation}
\begin{equation}
    u = U + Z_U +C_U(U-B) + k_U X
\end{equation}

\begin{figure}[th]
    \centering
    \includegraphics[width= 8.8cm, height=8cm]{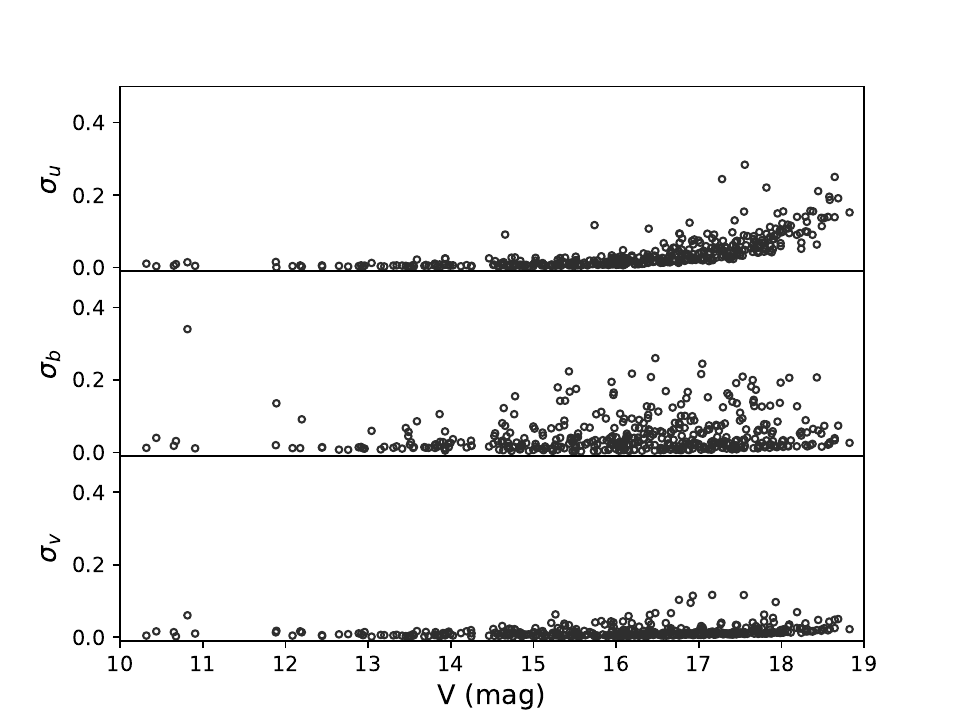}
    \caption{Photometric errors in different filters against {\textit{V}} magnitude.}
    \label{fig:v_band&error_new}
\end{figure}

Where $u,b,v$ denote the instrumental magnitudes, $U$, $B$, $V$ is the standard magnitudes, $X$ is airmass, and $k_i$ are the extinction coefficients. The calculated color coefficients (C) and zero points (Z) for different filters are listed in Table \ref{tab: color&zeropoint}. The error in color coefficients and zero points are $~0.01-0.04$ mag. The internal errors in each filter derived from DAOPHOT are plotted against $V$ magnitude in Fig. \ref{fig:v_band&error_new}. This figure shows that the average photometric error is $\leq$ 0.15 mag for the $B$ and $V$ filters at $V \sim$ $19^{th}$ mag, and it is $\leq$ 0.2 mag for the $U$ filter at $V \sim$ $19^{th}$ mag. Specifically, the error amount for the $V$ filter is 0.08 mag at 16$^{th}$ and 0.1 mag at 19$^{th}$ mag. 

\begin{table}
\caption{Derived Standardization coefficients and its errors.}
\begin{center}
\begin{tabular}{|c|c|c|}
\hline
 Filters &Color Coeff(C)  & Zeropoints(Z) \\
\hline
 $V$   &  $-$0.001 $\pm$ 0.045    & 0.75 $\pm$ 0.04  \\
\hline
 $B$ & 0.105$\pm$ 0.039 & 1.18$\pm$ 0.03\\
\hline
 $U$ & $-$0.166 $\pm$ 0.017 & 3.51 $\pm$ 0.01   \\
\hline
\end{tabular}
\end{center}
\label{tab: color&zeropoint}
\end{table}

\begin{figure*}[th]
    \centering
    \includegraphics[width= 12.5cm, height=7.5cm]{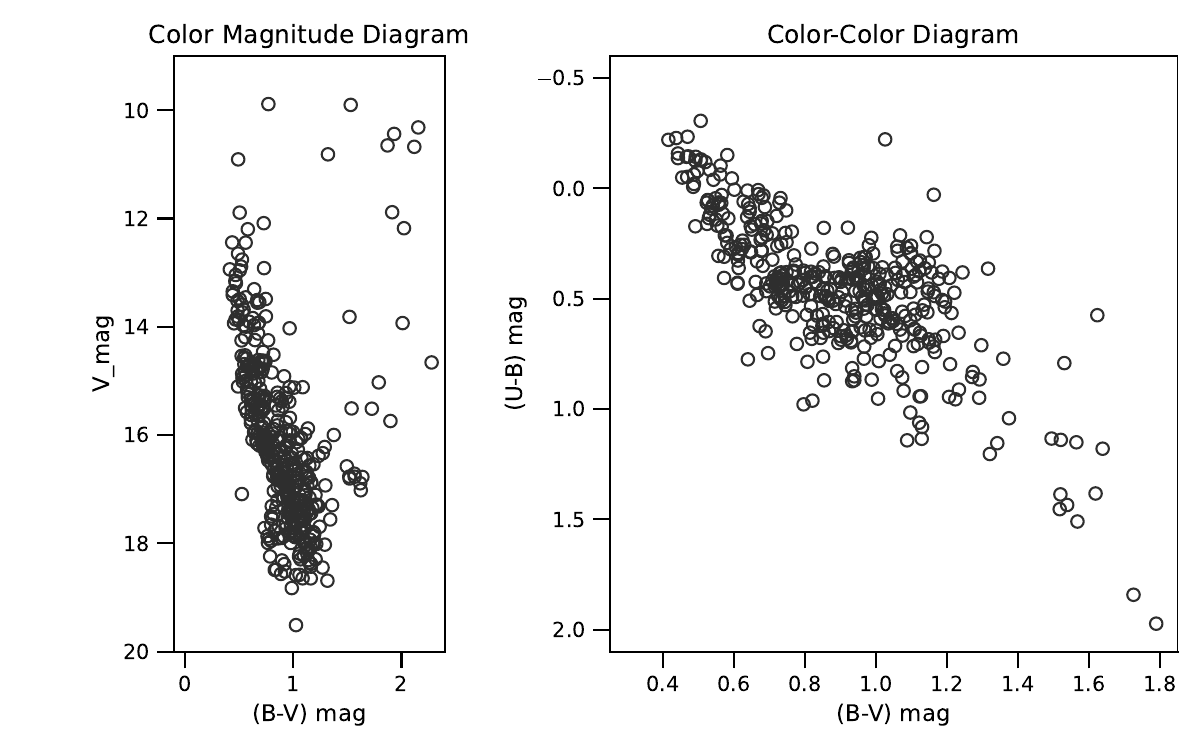}
    \caption {The $V/(B-V)$ color-magnitude and $(U-B)/(B-V)$ color-color diagrams of the cluster NGC 2345. A total of 426 stars observed in the present analysis are used for these diagrams.}
    \label{fig:1new_426_all_stars_cmd.pdf}
\end{figure*}

We converted the CCD X and Y pixel position of stars into right ascension (RA) and declination (DEC) of J2000 using the CCMAP and CCTRAN tasks provided in the IRAF. The resulting RA and Dec have standard deviations of $\sim 0.1$ arcsec.\\ 
We plotted the color-magnitude and color-color diagram, as shown in Fig. \ref{fig:1new_426_all_stars_cmd.pdf}, using 426 stars from HCT observation. This figure shows the contamination of field stars that need to be removed from our sample to estimate parameters precisely.

\subsection{Archived data}

\subsubsection{Gaia DR3}
\begin{figure}[th]
    \centering
    \includegraphics[width=10cm]{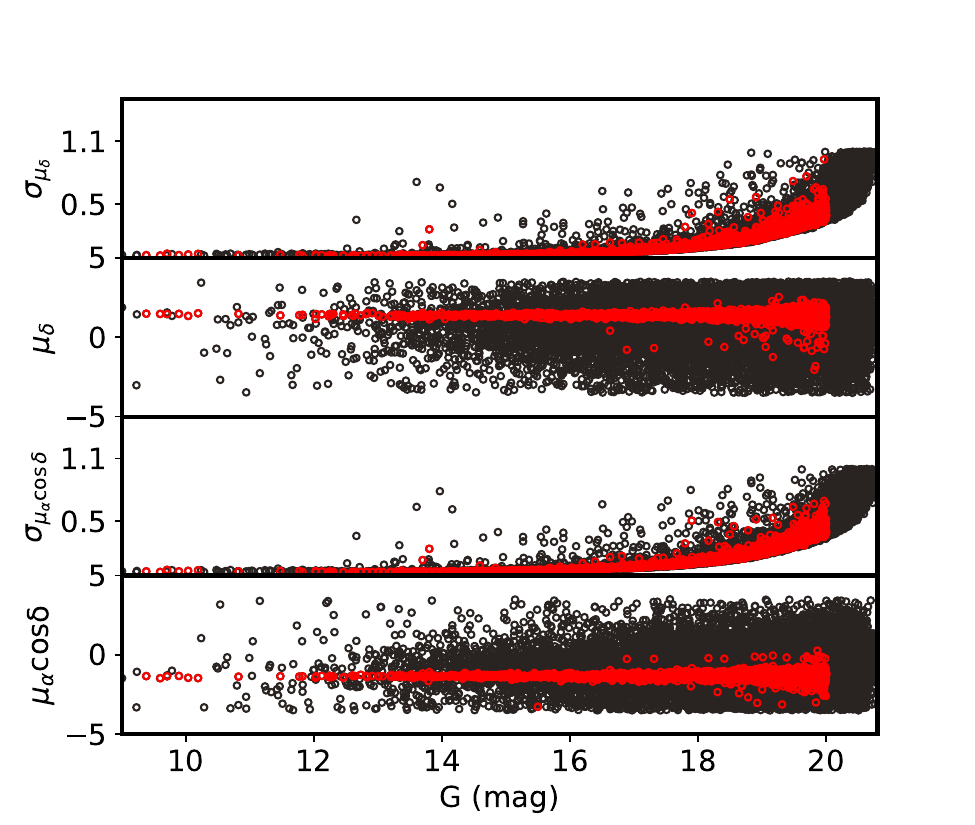}
    \caption{Plot of proper motions and their associated errors versus  {\textit{G}} magnitude. The red empty circle symbol represents stars having membership probability $\geq$70\%, and the black circle shows the field stars.}
    \label{fig:Plot_G&Pm}
\end{figure}

\emph{Gaia} DR3 \citep{vallenari2023gaia} is used for the astrometric study of NGC 2345 and determined structural parameters of the cluster. \emph{Gaia} DR3 provided celestial positions and {\textit{G}}-band magnitudes for a vast dataset of around 1.8 billion sources, with a magnitude measurement extending up to 21 mag. Additionally, \emph{Gaia} DR3 provides valuable parallax, proper motion, and color information ($G_{BP} – G_{RP}$) for a subset of this data set, specifically 1.5 billion sources. The uncertainties in parallax values are $\sim$ 0.02-0.03 milli arcsecond (mas) for sources at  {\textit{G}} $\leq$ 15 mag and $\sim$ 0.07 mas for sources with  {\textit{G}} $\sim$ 17 mag.  We collected data for NGC 2345 within a 25-arcminute radius, considering the previously estimated cluster core and tidal radii of 3.44 and 18.7 arcmin, respectively, by \cite{alonso2019comprehensive}. The cluster's proper motions and corresponding errors are graphically represented against  {\textit{G}} magnitude in Fig. \ref{fig:Plot_G&Pm}. The uncertainties in the corresponding proper motion components are $\sim$ 0.01-0.02 mas $yr^{-1}$ (for  {\textit{G}} $\leq$ 15 mag), $\sim$0.05 mas $yr^{-1}$ (for  {\textit{G}} $\sim$ 17 mag), $\sim$0.4 mas $yr^{-1}$ (for  {\textit{G}} $\sim$ 20 mag) and $\sim$1.4 mas $yr^{-1}$ (for $ G$ $\sim$ 21 mag).
\\

\subsubsection{2MASS}

This study used 2MASS (Two Micron All-Sky Survey) data for the cluster NGC 2345. This data set has been collected via the two highly automated 1.3-m telescopes, one at Mt Hopkins, Arizona (AZ), USA, and the other at CTIO, Chile, with three 3-channel cameras (256x256 array of HgCdTe detectors). The 2MASS database comprises photometric data in the near-infrared $J$, $H$, and $K$ bands, reaching limiting magnitudes of 15.8, 15.1, and 14.3, respectively. This data is available with a signal-to-noise ratio (S/N) greater than 10. We performed a cross-match of our dataset with 2MASS data using the Topcat \footnote[3]{https://www.star.bris.ac.uk/~mbt/topcat/} software.

\subsubsection{APASS}

The American Association of Variable Star Observers (AAVSO) Photometric All-Sky Survey (APASS) is organized in five filters:  {\textit{B}},  {\textit{V}} (Landolt), and $g', r,' i'$ providing stars with $V$ magnitude ranges from 7 to 17 mag \citep{henden2014apass}. Their latest catalog, DR9, covers almost 99\% sky. For NGC 2345, we downloaded data from APASS catalogue\footnote[4]{https://vizier.cds.unistra.fr/viz-bin/VizieR?-source=II/336}.

\subsection{Comparison with previous photometry}
\begin{figure}[th]
    \centering
    \includegraphics[width=9cm]{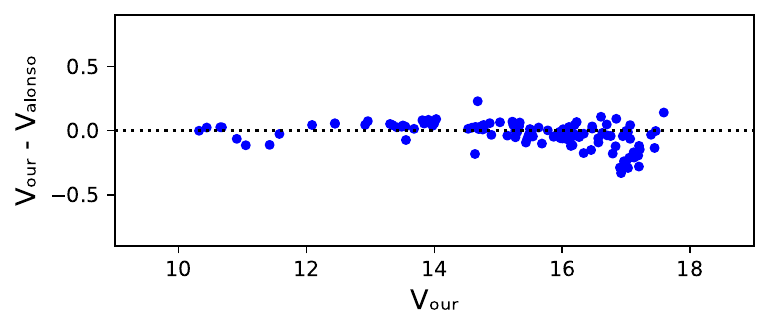}
    \caption{The blue dots on the graph represent the differences between our photometry and Alonso's against V magnitude. The dotted line indicates zero difference.}
    \label{fig:match-alonso}
\end{figure}

The CCD \emph{ubvy} photometry down to $V$ $\sim$ 18.0 for the open cluster NGC 2345 has been discussed by \cite{alonso2019comprehensive}. We have performed a cross-identification of stars in the two catalogs, considering stars to be correctly matched when the positional difference is within one arc-second. Based on these criteria, we have successfully identified 203 common stars. Figure \ref{fig:match-alonso} compares $V$ magnitudes between the two catalogs. In table \ref{tab:match-alonso} second column, we list the difference between our $V$ magnitude and that of \citep{alonso2019comprehensive}. This indicates that our $V$ magnitudes measurements agree with those provided in \cite{alonso2019comprehensive}.\\

\begin{table}
\caption{Comparision of our Photometry with Alonso-Santiago et al. (2019) and APASS catalog. The standard deviation in the difference for each magnitude bin is also given in parentheses.}
\begin{center}
\begin{tabular}{|c|r|c|c|}
\hline
 $V$ & $\Delta V$ (Alonso) & $\Delta$ $V$ (APASS) & $\Delta(B-V)$ (APASS) \\
\hline
 10-11 & 0.00(0.03)  &  0.01(0.04)   & 0.15(0.14)  \\
\hline
 11-12 & $-$0.08(0.04) &  0.06(0.01) & 0.02(0.01) \\
\hline
 12-13 & 0.04(0.01) &  0.01(0.01) & 0.00(0.01)  \\
\hline
 13-14 & $-$0.15(0.10) & 0.05(0.10)    & 0.02(0.03) \\
\hline
 14-15 & 0.02(0.16) &  0.16(0.30)  & 0.04(0.04) \\
\hline
 15-16 & $-$0.07(0.28) & 0.08(0.30)  & 0.01(0.13) \\
\hline
 16-17 & 0.12(0.56) & -- & --\\
\hline
 17-18 & 0.20(0.96) & -- & -- \\
 \hline
\end{tabular}
\end{center}
\label{tab:match-alonso}
\end{table}

\subsection{Comparison with APASS photometry}
\begin{figure}[th]
    \centering
    \includegraphics[width=9cm]{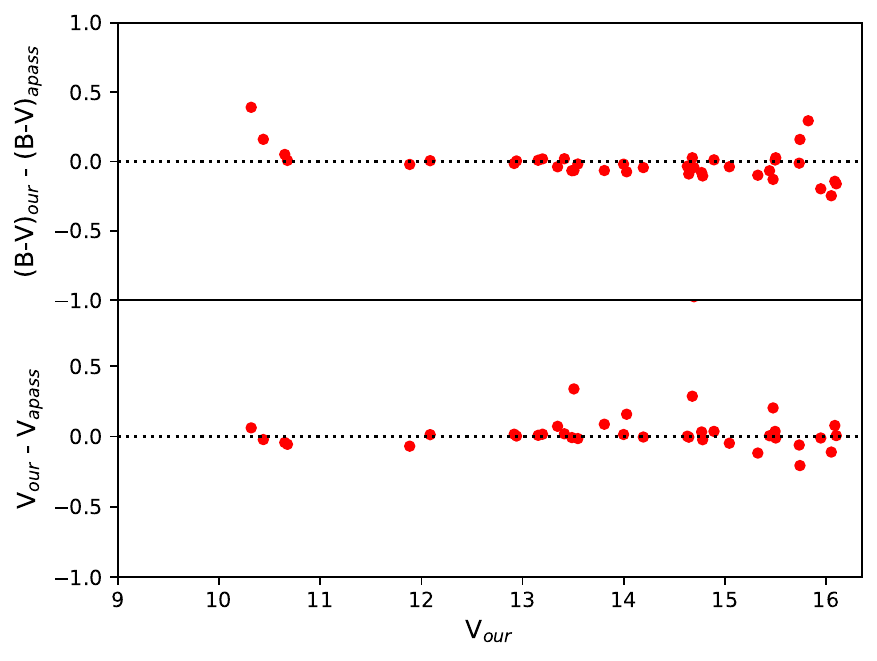}
    \caption{Differences between measurements presented in the APASS catalog and this study for $V$ magnitude and $(B-V)$ colors. The dotted line on the plot represents zero difference.}
    \label{fig:match_apass}
\end{figure}

We conducted a cross-match between the current and APASS catalogs to compare photometry data. We considered a maximum positional difference of 1.5 arcseconds during the matching process. As a result, we identified 52 common stars in both catalogs. A comparison of $V$ magnitudes and $(B - V)$ color between the two catalogs is plotted against $V$ magnitude and shown in Fig \ref{fig:match_apass}. The results, including the mean differences and standard deviations in each magnitude bin, are listed in Table \ref{tab:match-alonso}. The results of this comparison indicate a good agreement between the $V$ and $(B - V)$ measurements in our catalog and those provided in the APASS catalog.

\section{Membership probability}
\label{sec:Membership probability}
\begin{figure*}[th]
    \centering
    \includegraphics[width=15cm,height=14.0cm]{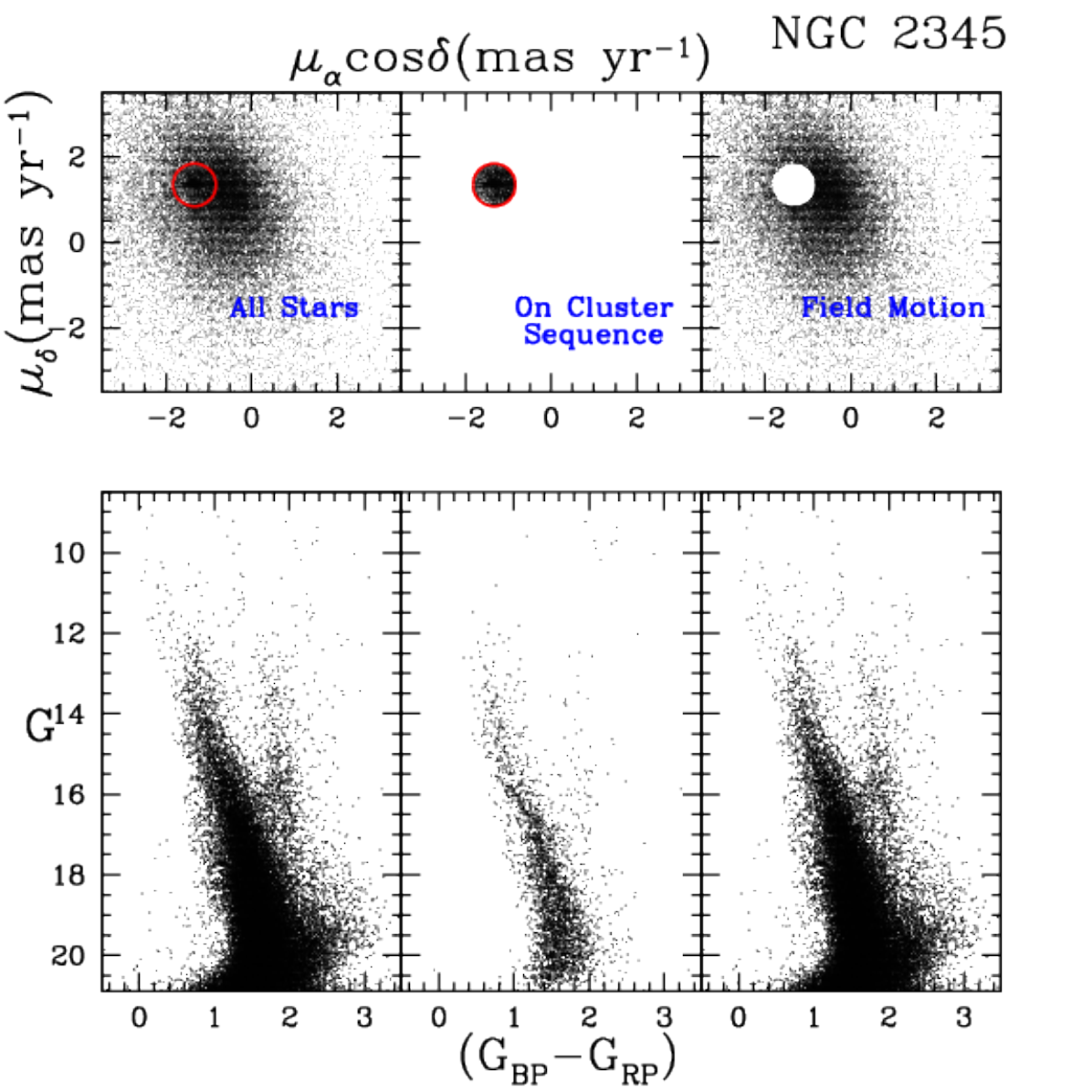}
    \caption{This figure shows the initial cluster member separation method using  Gaia proper motion for the cluster. The top panel shows the vector point diagram and the bottom panels show the respective Gaia color-magnitude diagrams for the total stars, cluster members, and field stars. In the top panel, draw a circle with a radius of 0.5 mas yr$^{-1}$  inside the cluster field, representing the member stars. }
    \label{fig:vpd}
\end{figure*}

\begin{figure}[th]
    \centering
    \includegraphics[width=8cm]{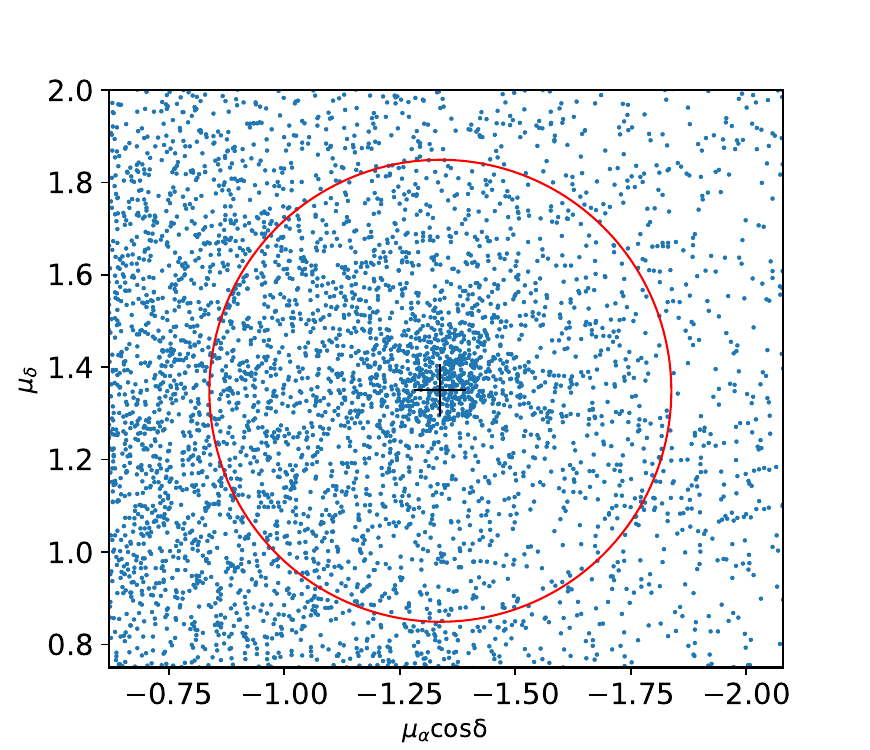}
    \caption{ A zoom view of the proper motion distribution of stars within a red circular area with a radius of 0.5 mas $yr^{-1}$. The plus sign denotes the proper motion center.}
    \label{fig:proper_motion_dist}
\end{figure}

Most open clusters are found within the Galactic disk, contaminating them by numerous foreground and background stars. It is essential to distinguish between cluster members and non-members to obtain more accurate fundamental parameters for the cluster \citep{dias2018astrometric}. We used the membership determination method described in \cite{balaguer1998determination} for the cluster NGC 2345. Many authors used this method in previous studies (\cite{yadav2013proper}; \cite{bisht2021detailed,bisht2021deep,bisht2022comprehensive,bisht2022deep11}; \cite{sharma2020disintegrating}).
Proper motion (PM) distribution is the unique identity of the stars or the excellent tool to separate the cluster members. In the PMs plane, cluster members are more concentrated than the field stars because member stars have similar properties. This method is purely based on the PM distribution of the stars. The \emph{Gaia} DR3 catalog offers unprecedented astrometric precision, enhancing reliability when determining cluster membership through kinematic data. As a result, we employed data from the \emph{Gaia} DR3 archive for our kinematic analysis, membership determination, and distance calculations of the cluster. We considered probable cluster members selected from vector point diagram (VPD) and color-magnitude diagram (CMD) to estimate the mean proper motion, as shown in Fig. \ref{fig:vpd}. 
 A radius 0.5 mas $yr^{-1}$ circle around the center of the proper motion distribution is drawn and assumed to be cluster members. The remaining sources are considered field stars. We have displayed the CMD of the most probable members in the lower-middle panel, where the cluster's main sequence appears well-defined for NGC 2345. The VPD based on the PMs in right ascension ($\mu_{\alpha^\ast}$) and declination ($\mu_{\delta}$) plane zoomed view is shown in Fig \ref{fig:proper_motion_dist}, where $\mu_{\alpha^\ast} =\mu_{\alpha}cos(\delta)$. 

In our selection method, we picked a star as the most probable member if it lies within the used radius in VPD, having a PM error of $\le 0.5 ~$ mas yr$^{-1}$ and has a parallax within 3$\sigma$
from the mean parallax of the cluster. The proper motion of the stars in the cluster region reaches its maximum value at ($\mu^{*}_{\alpha}$, $\mu_{\delta}$ ) = (-1.34, 1.35) mas $yr^{-1}$.  We have used the approach discussed in \cite{bisht2022comprehensive} to calculate the membership probability of the stars in the cluster region. We found the dispersion in PMs as 0.08 mas $yr^{-1}$ by using cluster distance as 2.78 kpc and the radial velocity dispersion of 1 km $s^{-1}$ for open clusters \citep{girard1989relative}. For field members, we have estimated ($\mu_{xf}$ , $\mu_{yf}$ ) = (1.3,-0.3) mas $yr^{-1}$ and ($\sigma_{xf}$, $\sigma_{yf}$) = (3.1,2.9) mas $yr^{-1}$ . These values are used to construct the frequency distributions of the cluster stars ($\phi_{\nu c}$) and field stars ($\phi_{\nu f}$) using the below equations described in \cite{yadav2013proper}, and \cite{bisht2020comprehensive}-

\begin{multline*}
\phi_{c}^{\nu} = \frac{1}{2\pi\sqrt{(\sigma^{2}_{c}+\epsilon^{2}_{xi})(\sigma^{2}_{c}+\epsilon^{2}_{yi})}} \\
\times \exp \Biggl\{-\frac{1}{2} \Biggl[\frac{(\mu_{xi}-\mu_{xc})^{2}}{\sigma_{c}^{2}+\epsilon^{2}_{xi}} + \frac{(\mu_{yi}-\mu_{yc})^{2}}{\sigma_{c}^{2}+\epsilon^{2}_{yi}}\Biggr]\Biggr\}
\end{multline*}

\begin{multline*}
\phi_{f}^{\nu} = \frac{1}{2\pi\sqrt{(1-\gamma^{2})}\sqrt{(\sigma^{2}_{xf}+\epsilon^{2}_{xi})(\sigma^{2}_{yf} +\epsilon^{2}_{yi})}}\\  \times \exp \Biggr\{-\frac{1}{2(1-\gamma^{2})} \Biggr[\frac{(\mu_{xi}-\mu_{xf})^{2}}{\sigma_{xf}^{2}+\epsilon^{2}_{xi}} \\ - \frac{2\gamma(\mu_{xi}-\mu_{xf})(\mu_{yi}-\mu_{yf})}{\sqrt{(\sigma_{xf}^{2}+\epsilon^{2}_{xi})(\sigma_{yf}^{2}+\epsilon^{2}_{yi})}} - \frac{(\mu_{yi}-\mu_{yf})^{2}}{\sigma_{yf}^{2}+\epsilon^{2}_{yi}}\Biggr]\Biggr\}
\end{multline*}

Where ($\mu_{xi}$, $\mu_{yi}$) represents the proper motions (PMs) of the $i^{th}$ star, and ($\epsilon_{xi}$, $\epsilon_{yi}$) are the corresponding errors in the PMs. The coordinates ($\mu_{xc}$, $\mu_{yc}$) designate the PM center of the cluster, while ($\mu_{xf}$, $\mu_{yf}$) symbolize the PM center coordinates for field stars. The intrinsic proper motion dispersion for cluster members is represented by $\sigma_c$, whereas $\sigma_{xf}$ and $\sigma_{yf}$ show the intrinsic proper motion dispersions for field stars. The correlation coefficient $\gamma$ is calculated as:

\begin{equation}
\begin{split}
    \gamma = \frac{(\mu_{xi}-\mu_{xf})(\mu_{yi}-\mu_{yf})}{(\sigma_{xf}\sigma_{yf})}
\end{split}
\end{equation}

\begin{figure*}[th]
    \centering
    \includegraphics[width=8cm,height=8cm]{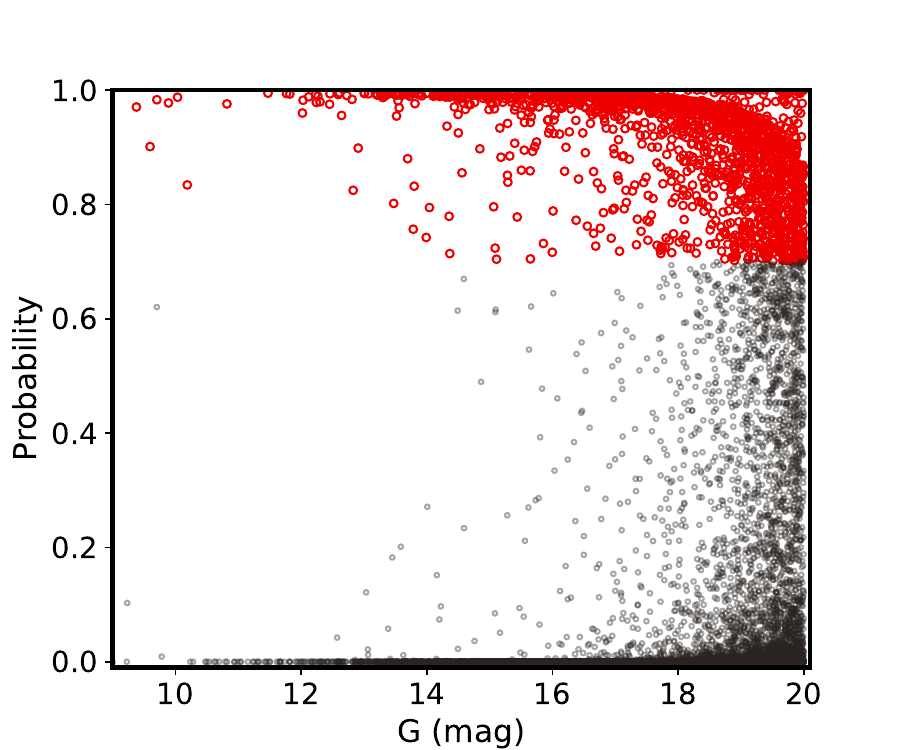}
    \includegraphics[width=7cm,height=9.3cm]{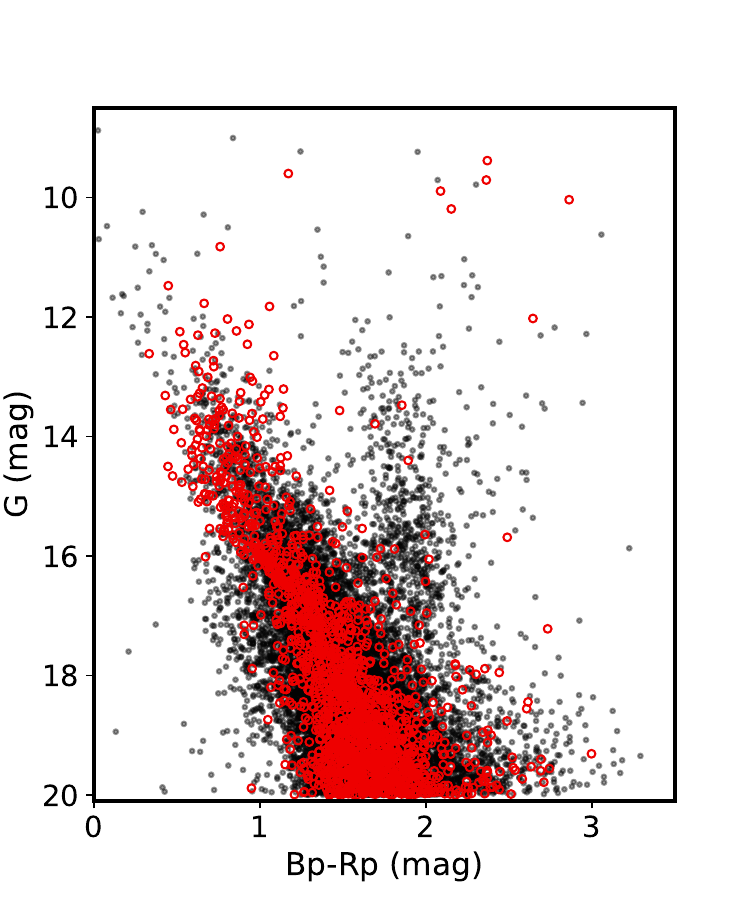}
    \caption{The  \emph{Gaia} membership probability is plotted as a function of $G$ magnitude, with cluster members indicated by red empty circles with membership probability higher than 70$\%$. The distribution of stars in the Color-Magnitude Diagram (CMD) is shown in the right panel.}
    \label{fig:Gmag&prob}
\end{figure*}

Finally, the membership probability for the $i^{th}$ star is eventually determined through the utilization of the following equation:

\begin{equation}
    P_{\mu}(i) = \frac{n_{c}\times \phi^{\nu}_{c}(i)}{n_{c}\times \phi^{\nu}_{c}(i) + n_{f} \times \phi^{\nu}_{f}(i)}
\end{equation}
$n_{c}$ and $n_{f}$ are the normalized number of probable cluster members and field members, respectively. A plot of membership probability with  {\textit{G}} mag is shown in Fig. \ref{fig:Gmag&prob}. We have found 1732 stars with probability $\geq 70\%$, shown by red symbols in this Figure. There is a clear separation between cluster stars and the field stars in the brighter end. At the fainter end ($G \sim$ 20 mag), the separation between the stars gradually decreases. Most stars exhibiting high membership probability display a closely clustered distribution in the plot. We have used only those identified cluster members for further analysis in this paper.

\section{Structural Parameters of the Cluster}\label{sec:Structural Parameters of the Cluster}

\begin{figure*}[th]
    \centering
    \includegraphics[width= 8cm,height=7.0cm]{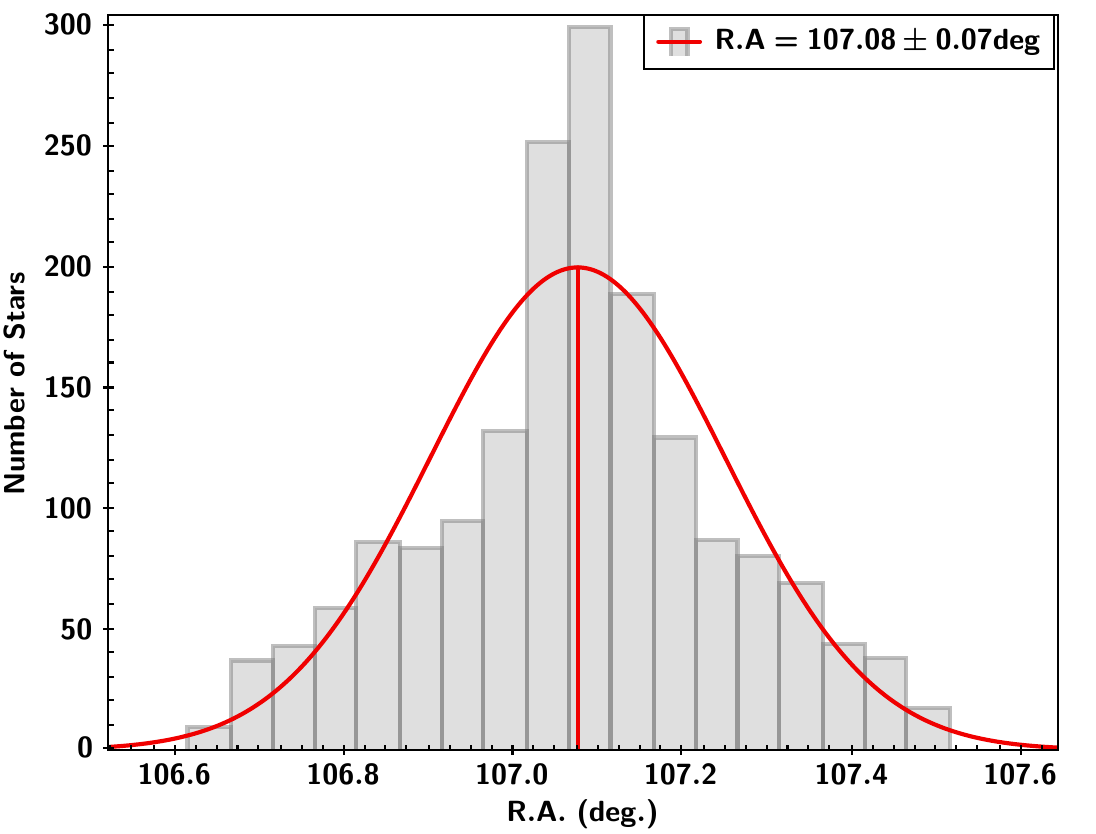}
    \includegraphics[width= 8cm,height=7.0cm]{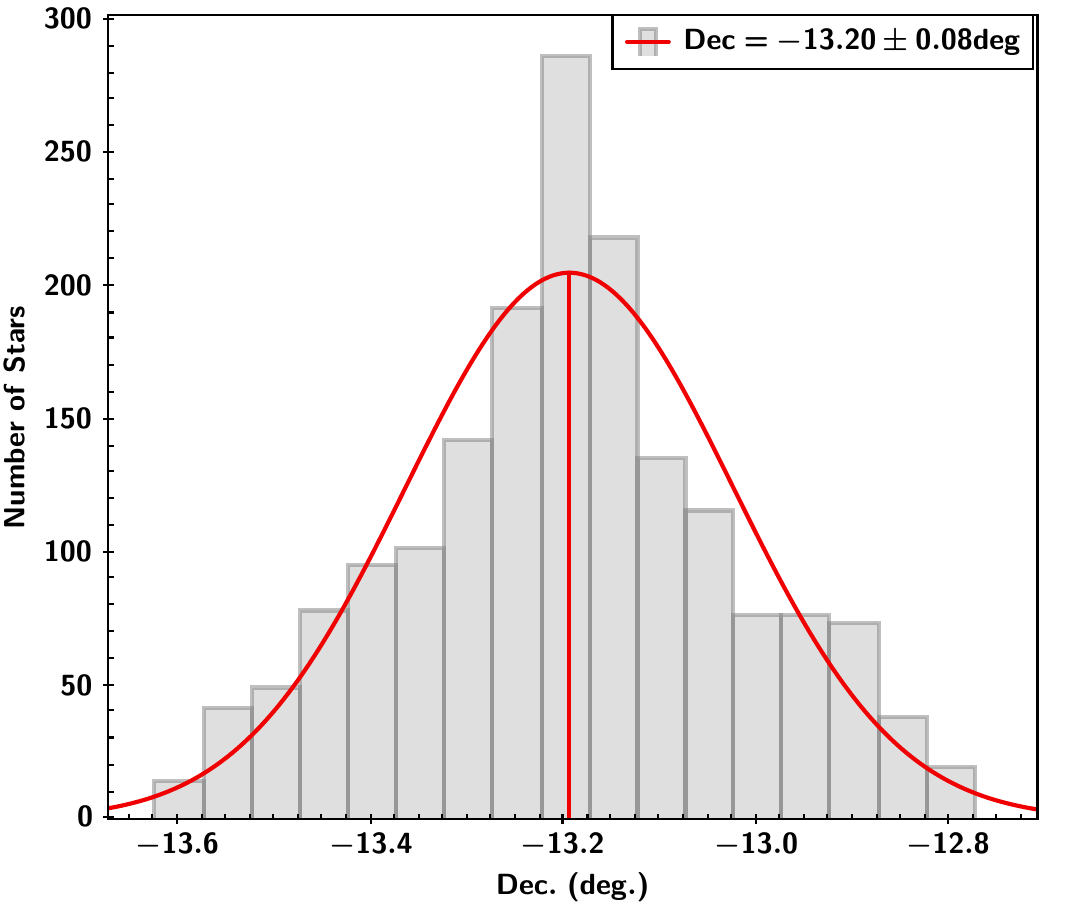}
    \caption{Stellar count profiles across the cluster region have been analyzed using Gaussian fits. The center of symmetry about the peaks of Right Ascension and Declination is the position of the cluster’s center.}
    \label{fig:ra&dec_hist}
\end{figure*}

Generally, open clusters consist of thinly dispersed and loosely bound stars, yet they exhibit the highest stellar density at their center. To estimate the center coordinate of the cluster NGC 2345, the Gaussian curve fitting provides the central coordinate as $\alpha$=107.08$\pm$0.07 deg and $\delta = -13.20 \pm 0.08$ deg, which is shown in Fig. \ref{fig:ra&dec_hist}. These values match the values given by \cite{cantat2018gaia}.

The cluster radius $(r_{cl})$ is defined as the distance from the cluster center at which the average cluster contribution becomes negligible compared to the background stellar field. We utilize the star's spatial surface density profile to estimate the cluster radius and assess the extent of field-star contamination. We achieve this by considering concentric circular regions around the estimated cluster center. Within each annular ring, typically with a width of approximately 70 arc-seconds, we count the stars and then divide this count by the respective areas of the annular rings to obtain the number density $\rho_{i} = N_{i}/A_{i}$, where $N_{i}$ is the number of stars and $A_{i}$ is the area of the $i^{th}$ zone. To derive the structural parameters, we fitted a surface density profile by \cite{king1962structure} to the radial distribution of stars. This must be done using a non-linear least squares fitting, which uses the errors as weight. 
\begin{figure}[th]
    \centering
    \includegraphics[width=9cm]{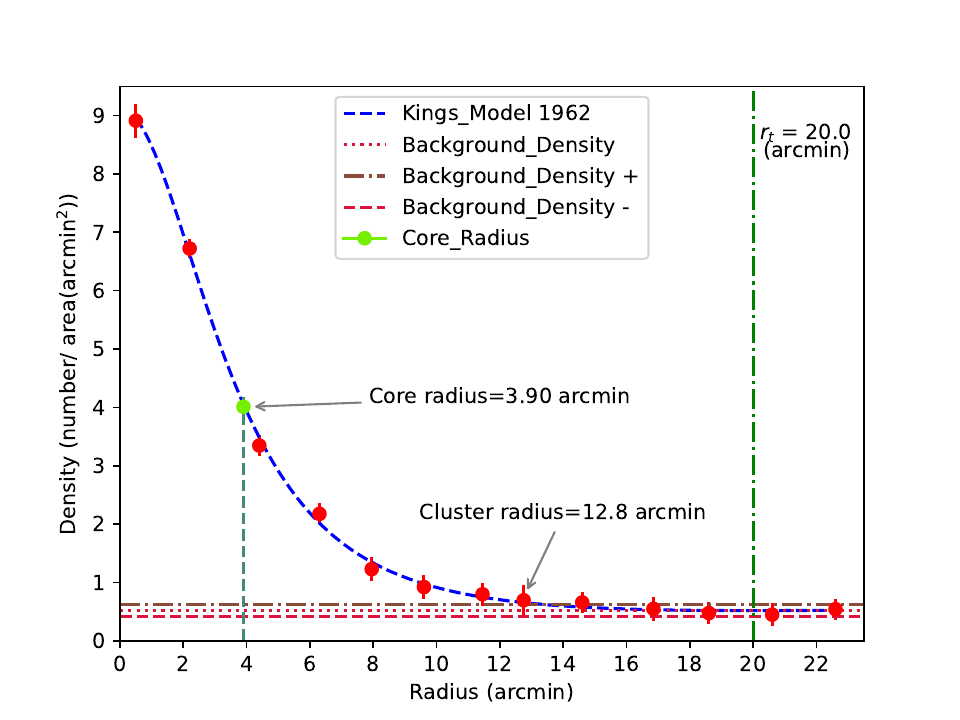}
    \caption{ The radial density proﬁle for the cluster. The best-ﬁt blue curve represents the King (1962) proﬁle, from which we obtained both the core ($r_{c}$ ) and the tidal ($r_{t}$ ) radius. Their positions are indicated with vertical dashed lines. The horizontal lines represent the star's background density and their corresponding errors.}
    \label{fig:rdp}
\end{figure}
Fig \ref{fig:rdp} displays the best-fit solution for the density distribution and its associated uncertainties. For cluster, it decreases and flattens around $r_{cl} = 12.8 \pm 0.6$ arcmin and begins to merge with the field star density. Therefore, we consider 12.8 arcmin as the cluster radius. A smooth dashed line represents a fit King's (1962) profile:\\

\begin{equation}
    f(r) = f_{0} \Biggl[\frac{1}{\sqrt{1 + (r/r_c)^{2}}} - \frac{1}{\sqrt{1+(r_t/r_c)^{2}}}\Biggr]^{2} + f_{b}
\end{equation}

where $f_{b}$ is the background density, $f_{0}$ is the central density, $r_{c}$ is core radius and $r_{t}$ is the tidal radius. The core radius of a cluster can be defined as the distance from the cluster center at which the stellar density becomes half of the density at the cluster's center. The central density, background density, core radius, and tidal radius values are in Table \ref{tab: kings model parameters}.

\begin{table}[th]
\centering
\caption{Structural Parameters of the Cluster under Study}
\begin{tabular}{|c|c|}
\hline
 Central Density$(f_{0})$ & 10.12 $\pm$ 0.98  number/arcmin$^{2}$  \\
 Background Density$(f_{b})$ & 0.51 $\pm$ 0.04  number/arcmin$^{2}$ \\
 Core Radius$(r_{c})$  &     3.9  $\pm$ 0.2 arcmin    \\
 Tidal Radius$(r_{t})$  &   20.0  $\pm$  2.1  arcmin  \\
 Cluster Radius$(r_{cl})$ & 12.8  $\pm$  0.6  arcmin \\
\hline
\end{tabular}
\label{tab: kings model parameters}
\end{table}

\section{Reddening of NGC 2345}\label{sec: Reddening NGC 2345}
The plots of two-color diagrams (TCDs) for various pairs of colors serve as valuable tools for estimating interstellar reddening and gaining insights into the properties of the extinction law in the direction of the clusters.\\

\subsection{$(U-B)/(B-V)$ color–color diagram}

\begin{figure}[th]
    \centering
    \includegraphics[width=9.0cm,height=8cm]{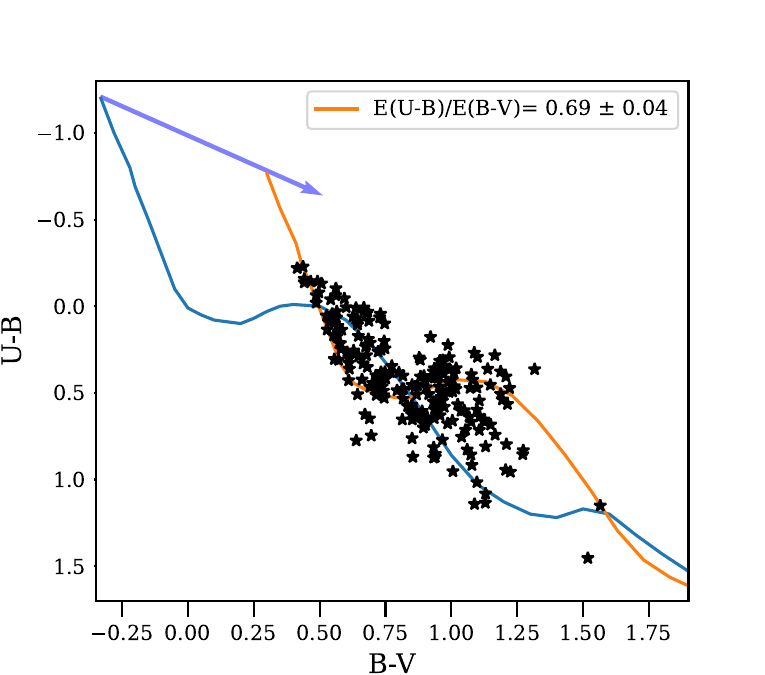}
    \caption{The $(U-B)$ vs $(B-V)$ color-color diagram for NGC 2345. We have used only cluster members with membership probability higher than 70$\%$. The arrow indicates the reddening vector.}
    \label{fig:U-B&B-V.pdf}
\end{figure}

Knowledge of reddening is crucial for determining the intrinsic properties of cluster stars. In cases where spectroscopic observations are unavailable, we can rely on color-color diagrams like $(B-V)$ and $(U-B)$ to estimate the reddening of clusters, as demonstrated by \cite{becker1954drei}. To estimate the interstellar extinction towards the clusters, we created $(U-B)$ versus $(B-V)$ diagram utilizing 233 identified cluster members, cross-referenced with Gaia data as illustrated in Fig \ref{fig:U-B&B-V.pdf}.\\

We acquired a reddening vector slope of $\frac{E(U-B)}{E(B-V)}$ and aligned it with the intrinsic Zero-Age Main Sequence (ZAMS) of 0.01 metallicity based on \cite{schmid1982landolt} for main-sequence stars. We adjusted the ZAMS to the brighter stars to account for varying color excesses. The optimal fit, associated with $E(B - V)$ = 0.63 $\pm$ 0.04 mag, is illustrated with the orange curve. The estimated $E(B - V)$ value shows good agreement with $\sim 0.66\pm 0.13 $ obtained by \cite{alonso2019comprehensive}.

\subsection{Total-to-selective extinction value}

Reddening is crucial for determining the fundamental parameters of clusters, such as age and distance. To ascertain the characteristics of the extinction law, it is essential to examine the two-color diagram(TCD). Photons emitted by the cluster stars that pass through the interstellar medium are absorbed and scattered by the medium's dust, gas, and molecular clouds. The normal Galactic reddening law frequently does not apply along the line of sight to clusters \citep{sneden1978infrared}.\\ 

\begin{figure}[th]
    \centering
    \includegraphics[width=10cm, height=7cm]{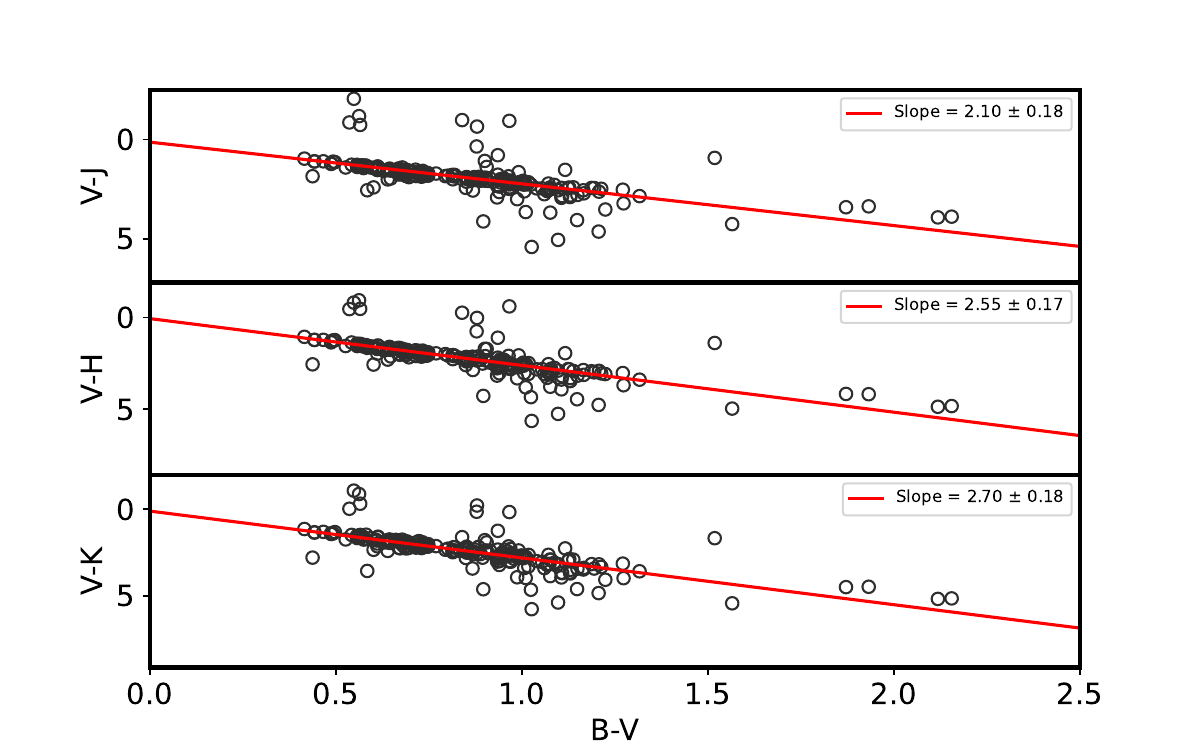}
    \caption{The ($V - \lambda)/(B - V)$ diagrams for member stars in the cluster NGC 2345. The continuous lines represent the slope determined through the least squares linear fit and are mentioned in each panel.}
    \label{fig:least_fit_tcd}
\end{figure}

\cite{chini1990large} suggested $(V - \lambda)$ vs. $(B - V)$ TCDs to examine the nature of the reddening law. Where $\lambda$ denotes the filter other than $V$ mag reddening law. Here, $\lambda$ denotes the magnitude with the filter at that wavelength. We have studied the reddening law for the cluster by creating TCDs in the three 2MASS filters, as shown in Fig \ref{fig:least_fit_tcd}. A linear dependence is observed among the stellar color values. The slope (m$_{cluster}$) for each TCD is calculated by least-square fitting as shown by the solid lines in Fig. \ref{fig:least_fit_tcd} and tabulated in Table \ref{tab:least_fit_tcd} with corresponding normal values. This table shows that the estimated color excess values align well with the expected normal values.  \\ 

\begin{table*}[th]
\caption{A comparison between the color-excess ratios observed in the direction of the cluster NGC 2345 and the standard values provided by Cardelli et al. (1989).}
\begin{center}
\begin{tabular}{|c|c|c|c|}
\hline
 NGC 2345 &  {\textit{(V-J)/(B-V)}} &  {\textit{(V-H)/(B-V)}} &  {\textit{(V-K)/(B-V)}} \\
\hline
 Observed ratio & 2.10 $\pm$ 0.18    &  2.55 $\pm$ 0.17    &  2.70 $\pm$ 0.18  \\
 Normal ratio   & 2.30        &  2.58    &  2.78            \\
\hline
\end{tabular}
\end{center}
\label{tab:least_fit_tcd}
\end{table*}

The extinction ($R_{cluster}$) towards the cluster is calculated by the following equation:\\

\begin{equation}
    R_{cluster} = \frac{m_{cluster}}{m_{normal}} \times R_{normal}
\end{equation}

Where R$_{normal}$ is the normal value of the total to selective extinction ratio, and m$_{cluster}$ and m$_{normal}$ are the estimated and normal slopes in the TCD, respectively. By assuming the value of R$_{normal}$ as 3.1, we calculated R$_{cluster}$ in the different passbands to be  $3.1\leq R \leq 3.4$, which is close to the normal value. Hence, the reddening law is similar to general interstellar medium within the cluster region.\\

\subsection{Interstellar reddening from JHK colors}
\begin{figure}[th]
    \centering
    \includegraphics[width=9.0cm,height=8cm]{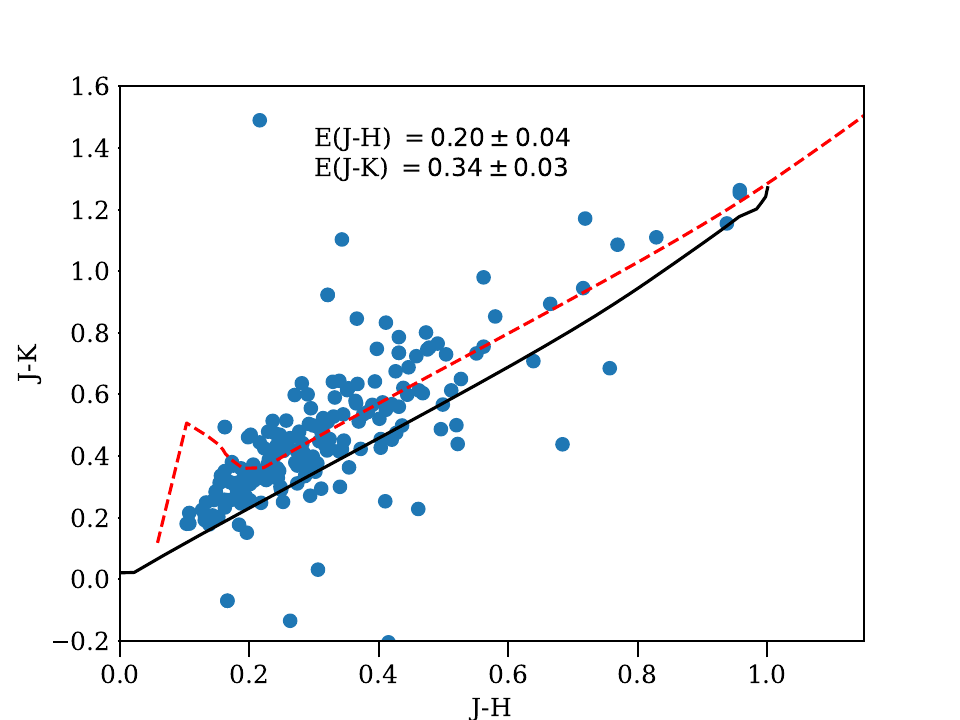}
    \caption{The plot of $(J - K)$ versus $(J - H)$ color-color diagram. Solid and dotted lines are the ZAMS taken from Caldwell et al. (1993).}
    \label{fig:J-k&J-H_plot}
\end{figure}

The near-IR photometry is important to understand the nature of interstellar extinction \citep{tapia1988interstellar}. The wavelength in the near-infrared (NIR) region is longer than in the visible spectrum. Therefore, it provides information about types of dust that have larger particle sizes compared to visible wavelength dust. As in the previous Section, we have utilized 2MASS data in the $JHK$ photometry bands to investigate the interstellar extinction law. The $(J - K)$ versus $(J - H)$ diagram is presented in Fig \ref{fig:J-k&J-H_plot}. \\

The Zero-Age Main Sequence (ZAMS) of 0.01 metallicity, depicted as the solid line, is adopted from \cite{caldwell1993statistical}. The visual fitting of ZAMS yields $E(J - H)$ = 0.20 $\pm$ 0.04 mag and $E(J - K)$ = 0.34 $\pm$ 0.03 mag. The ratio $\frac{E(J-H)}{E(J-K)}$ is obtained to be 0.59 $\pm$0.04, which shows a good agreement with the normal interstellar extinction value of 0.55 as suggested by \cite{cardelli1989relationship}. The reddening is calculated using the following equations \citep{fiorucci2003asiago}. \\

\begin{equation}
\begin{aligned} 
       E(J-H) &= 0.309 \times E(B-V)\\
       E(J-K) &= 0.480 \times E(B-V) 
\end{aligned}
\end{equation}

\begin{figure}[th]
    \centering
    \includegraphics[width=9.0cm,height=8cm]{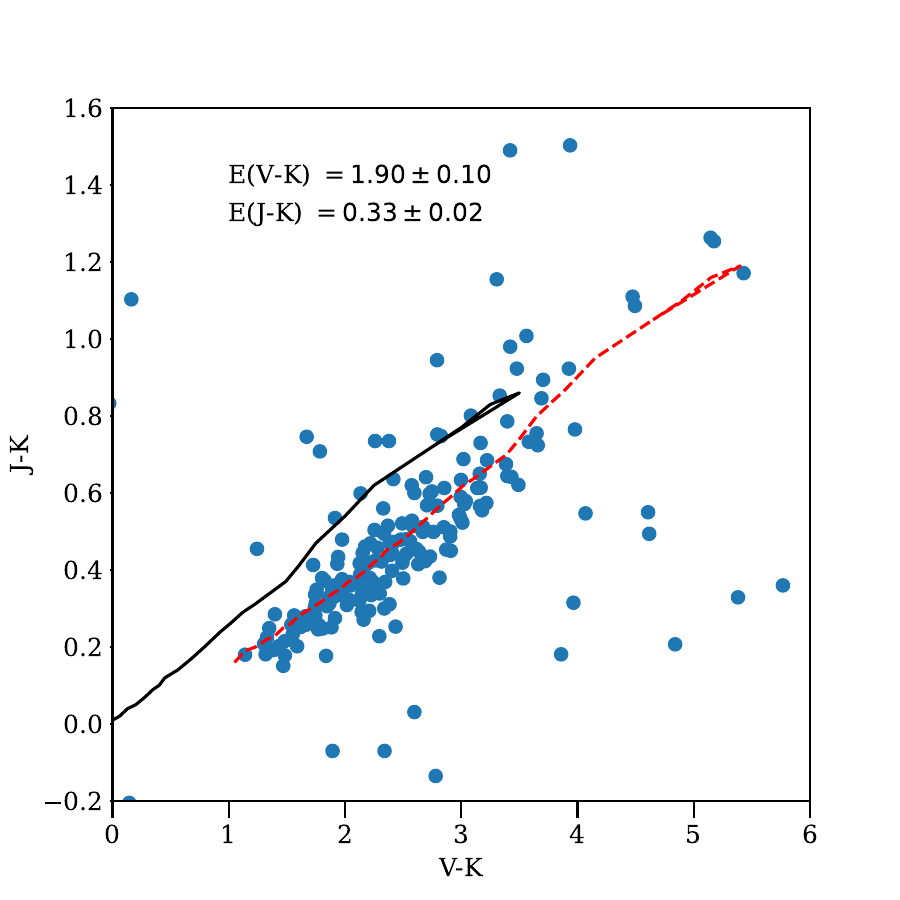}
    \caption{The plot of (V $-$ K) versus (J $-$ K) color-color diagram. Solid and dashed lines are the ZAMS taken from Caldwell et al. (1993).}
    \label{fig:V-K&J-K_plot}
\end{figure}

 We obtained the interstellar reddening of the cluster to be $0.68 \pm 0.10$, calculated by averaging the results from both equations, yielding individual values of $0.65 \pm 0.13$ and $0.71 \pm 0.06$. The $E(B-V)$ values derived from the near-IR TCDs reconfirm our finding of $E(B-V)$ value estimated from the optical ${E(U-B)}/{E(B-V)}$ TCDs. We have also calculated the value of $\frac{E(J-K)}{E(V-K)}= 0.17 \pm 0.06 $, which is in good agreement with the normal interstellar extinction value of 0.19 as given by \cite{cardelli1989relationship}, as shown in Fig \ref{fig:V-K&J-K_plot}.\\

\section{DISTANCE AND AGE OF THE CLUSTER }
\label{sec:DISTANCE AND AGE OF THE CLUSTER}
\subsection{Distance estimation using parallax}
\begin{figure}
    \centering
    \includegraphics[width=9cm]{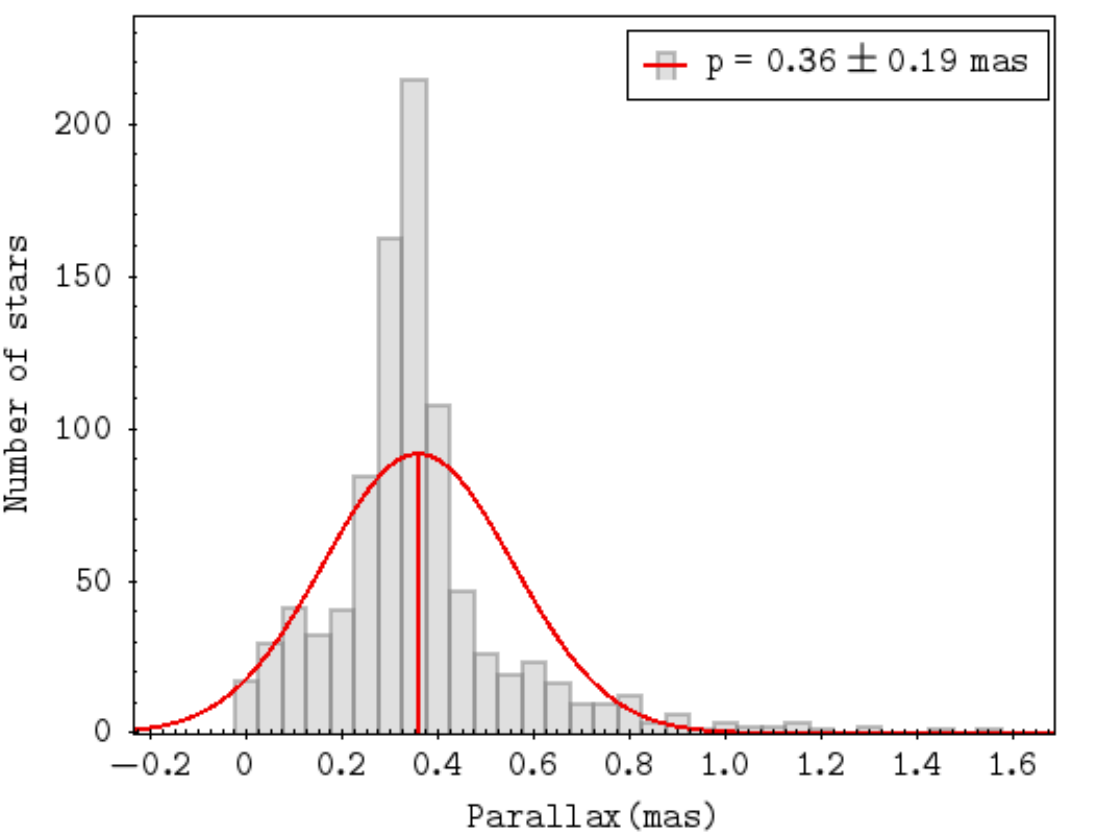}
    \caption{Plot of the histogram of parallax. The Gaussian function is fitted to the central bins and provides the mean value of parallax.}
    \label{fig:Parallax_correct_gaussian_fit_new}
\end{figure}

We employed the method proposed by \cite{luri2018gaia}, which involves estimating the distance of a cluster using its average parallax value. Because of the inherent errors in the parallax data from \emph{Gaia} DR3, we computed their weighted mean by considering the probable cluster member stars.  The parallax values from Gaia have negative values due to minimal angles hence indicators of large distances \citep{2016ApJ...833..119A}. Since the open clusters of the Galaxy are not that distant, we generated a histogram of the parallax values of cluster members, excluding stars with negative parallax values. We determined the mean parallax angle for the clusters by fitting Gaussian curves to the histograms. The plots for the histograms with Gaussian fits are presented in Fig \ref{fig:Parallax_correct_gaussian_fit_new}.\\

We estimated the mean parallax as 0.36 $\pm$ 0.19 mas and then computed cluster distance by reciprocating the mean parallax after applying a zero point offset of -0.021 mas to the mean parallax as suggested by \cite{groenewegen2021parallax} and obtained a distance of 2.77 $\pm$ 0.11 kpc. The parallax estimation by Gaia is associated with the corresponding errors; hence, the distance calculated by inverting the parallax can give the wrong estimation. \cite{bailer2015estimating} proposed a probabilistic analysis-based method for determining distances, which offers a more precise estimate than the distance obtained through parallax inversion. They used the Bayesian approach to calculate the distance of a star having a parallax and associated error measurements. For this, they tested different priors and concluded that exponentially decreasing space density prior and Milky Way priors are best among all. We have computed distances using the approach outlined in \cite{bailer2018estimating}. The calculated distance is 2.78 $\pm$ 0.78 kpc, slightly higher than $\sim$ 2.652 kpc obtained by \cite{cantat2020painting}.

\subsection{Age and distance from isochrone fitting}

\begin{figure*}[th]
    \centering
    \includegraphics[width=15cm,height=7cm]{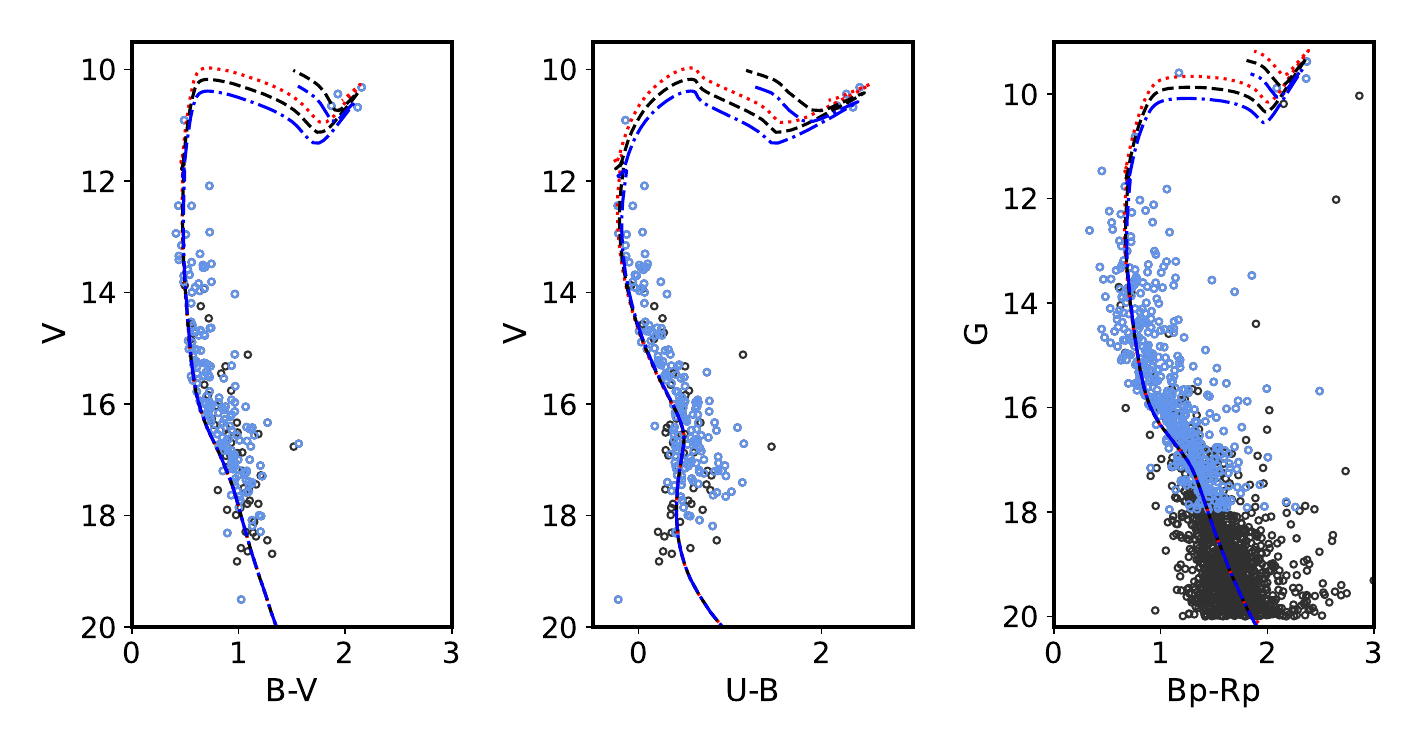}
    \includegraphics[width=15cm,height=7cm]{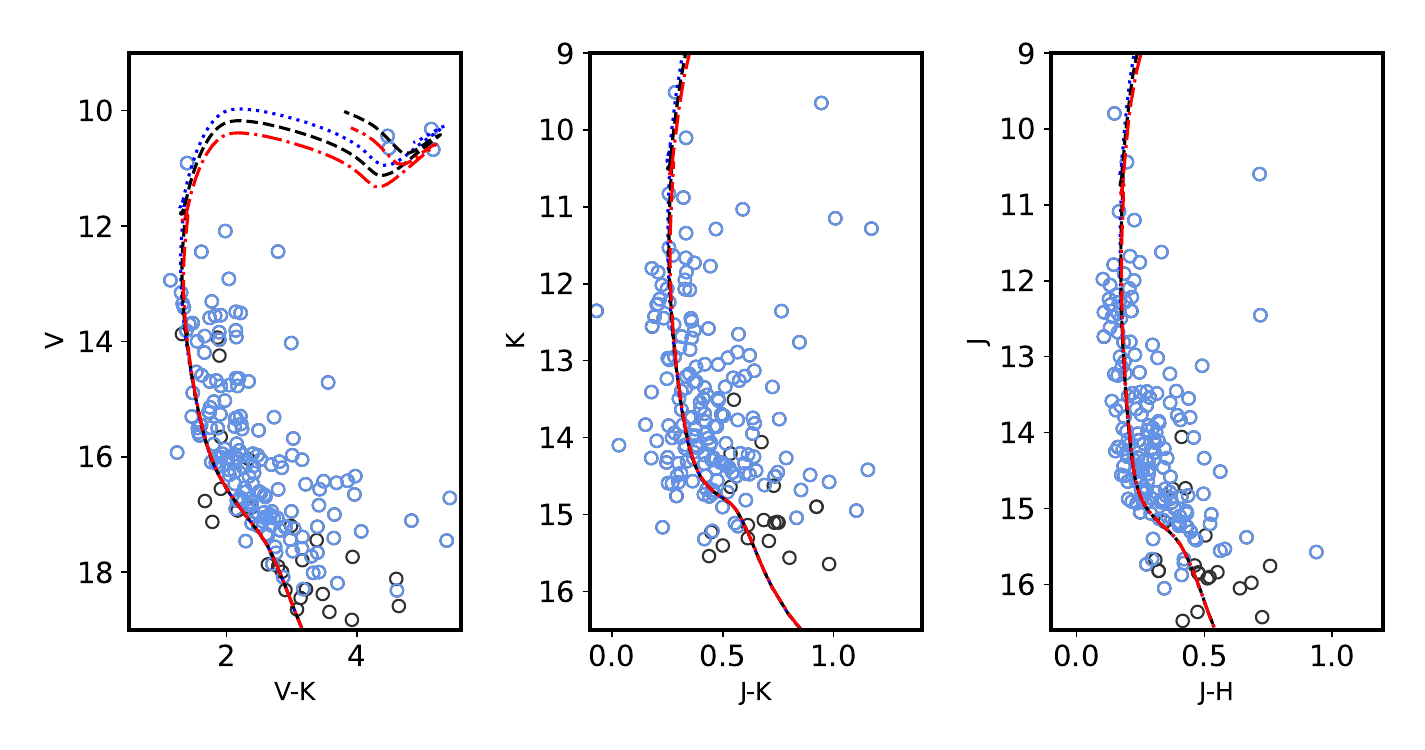}
    \caption{The Optical and near-IR color-magnitude diagram of the cluster under study. The curves are the isochrones of (log(age) = 7.75,7.8 and 7.85). These isochrones are taken from Marigo et al. (2017). Black circles are the probable cluster members identified in the present study, while the light blue circles represent the matched stars with the Cantat-Gaudin et al. (2020) catalog.}
    \label{fig:70prob_with_gaia_cmd_cantatgaudin_latest}
\end{figure*}

Color-magnitude diagrams (CMDs) illustrate the correlation between the absolute magnitudes of stars and their surface temperatures, typically identified by their color. CMDs are widely employed for studying star clusters and prove valuable for estimating their properties (\cite{kalirai2004interpreting}; \cite{sariya2021comprehensive}). We used the most probable member stars of NGC 2345 to generate the CMDs. Analyzing the morphology of CMDs helps identify the key features such as the main sequence, turnoff point, and giant members, ultimately leading to model-based mass, age, and distance estimations for each star \citep{bisht2019mass}. In this study, we simultaneously estimated the distance modulus and age by fitting the theoretical isochrones given by \cite{marigo2017new} to the \emph{UBV}, \emph{Gaia} and 2MASS-based CMDs of the cluster. To determine the distance modulus and age, distance of the cluster, we constructed V $\times$ $(B - V)$, $V$ $\times$ $(U - B)$, $V$ $\times$ $(V-K)$, $V$ $\times$ $(J - K)$, $J$ $\times$ $(J - H)$ and $G$ $\times$ ($G_{BP} - G_{RP}$) diagrams and visually fitted the isochrones \citep{marigo2017new} by only considering the probable members ($P \geq 0.7$). We attempted to fit numerous isochrones with varying metallicity and ages and determined the best fit at Z=0.01. We employed the magnitudes of stars in different filters to ensure a reliable determination of cluster properties via isochrone fitting. The $E(B-V)$ value is taken from section \ref{sec: Reddening NGC 2345} for the isochrones fitting procedure in $UBV$ filters. For \emph{Gaia} DR3 filters, we used relation $E(G_{BP} - G_{RP}) = 1.41 \times E(B-V)$ given by \cite{sun2021binary}, to convert $E(B-V)$ into E($G_{BP} - G_{RP}$). The calculated color excess value, E($G_{BP} - G_{RP}$)  for the cluster is 0.89 mag. Uncertainties in age were determined by utilizing both low and high-age isochrones, which were selected to best fit the observed scatter around the Main Sequence (MS) shown in the Fig \ref{fig:70prob_with_gaia_cmd_cantatgaudin_latest}.
\\

We superimposed theoretical isochrones of different ages (log(age) = 7.75,7.8 and 7.85 with z = 0.01) on the different CMDs to obtain distance modulus and ages. There are a few giants  present in this cluster. The fitted isochrones also pass through the giant stars, clearly illustrating the cluster's evolutionary path. In this way, we estimated the cluster's age to be 63 $\pm$ 8 Myr with a true distance modulus $(m-M)_0 = 12.0 \pm 0.2$ mag. The corresponding heliocentric distance, calculated through the distance modulus, is $2.51 \pm 0.12$ kpc. The distance calculated using distance modulus is compatible with \emph{Gaia} DR3 parallax distance measurement of the present study as well as the values reported by \cite{cantat2018gaia}, and \cite{alonso2019comprehensive} within a margin of errors. The $A_{V}$ value for cluster members was computed using the weighted mean method based on the CCD \emph{UBV} data.  Adopting $R_V =3.1$, the value $A_{V}$ for this cluster is 1.95 ± 0.12, which is consistent with the value of 1.91 reported by \cite{tsantaki2023search}. A comparative analysis of the fundamental properties of cluster members, as determined in this study, is presented alongside values found in the existing literature in Table \ref{tab:fundamental parameters}.\\
We have determined the galactocentric coordinates of the cluster as $X_{GAL}$ = -1.887 kpc, $Y_{GAL}$ = -1.994 kpc, and $Z_{GAL}$ = - 0.111 kpc.  The estimated galactocentric distance is $R_{GC}$=10.4 kpc. The above parameters are in fair agreement with the values obtained by \citep{cantat2020painting}.  Such information is essential for studying the cluster's environment, its motion within the galaxy, and its interaction with other galactic components. Additionally, galactocentric coordinates provide a foundation for broader astrophysical investigations, aiding in the contextual understanding of the cluster's role in the dynamic structure of the Milky Way.\\

\begin{table*}[th]
    \centering
    \begin{tabular}{|c|c|c|}
\hline
    Parameters & Numerical Values & Reference \\
\hline
    RA, DEC (deg)  & (107.08 $\pm$ 0.07, -13.20 $\pm$ 0.08)   & {\bf Present work}   \\
                   & (107.075, -13.199)  &  Cantat Gaudin et al. (2020)\\
                   & (107.085 $\pm$ 0.007, -13.197 $\pm$ 0.008) & Alonso et al. (2019)\\
                   & (107.075, -13.197)  & G Carraro et al. (2015)\\
    $\mu_{\alpha}cos\delta, \mu_{\delta}$ (mas $yr^{-1}$) & (-1.34 $\pm$ 0.20, 1.35 $\pm$ 0.21) & {\bf Present Work}\\
                   & (-1.33 $\pm$ 0.10, 1.34 $\pm$ 0.11) & R.Carrera et al. (2022) \\
                   & (-1.332, 1.340)  & Cantat Gaudin et al. (2020)\\
                   & (-1.36 $\pm$ 0.10, 1.33 $\pm$ 0.09) & Alonso et al. (2019)\\
    Cluster Radius (arcmin) & 12.8  & {\bf Present Work}\\
                            & 3.75   & G Carraro et al. (2015)\\
                            & 7.2    & Kharchenko et al. (2005)\\
                            & 5.25   & Moffat et al. (1974)\\
    Age (Myr)      & 63 $\pm$ 8    & {\bf Present Work}\\
                   & 56 $\pm$ 13 & Alonso et al. (2019)\\
                   & 79.4      & N Holanda et al. (2019)\\
                   & 63–70          & G Carraro et al. (2015)\\
                   & 77.4          & Kharchenko et al. (2005)\\
                   & 78.5          & Dias et al. (2002) \\
                   & 60            & Moffat et al. (1974)\\
    Mean Parallax (mas) & 0.36 $\pm$ 0.19 & {\bf Present Work} \\
                        & 0.35 $\pm$ 0.05 & R.Carrera et al. (2022)\\
                        & 0.348           & Cantat Gaudin et al. (2020)\\
                        & 0.35 $\pm$ 0.03 & Alonso et al. (2019)\\
    Distance (kpc)  & 2.78 $\pm$ 0.78 & {\bf Present Work}\\
                    & 2.662        & R.Carrera et al. (2022)\\
                    & 2.652        & Cantat Gaudin et al. (2020)\\
                    & 2.5  $\pm$ 0.2 & Alonso et al. (2019)\\
                    & 2.251        & Kharchenko et al. (2005)\\
                    & 1.750        & Moffat et al. (1974)\\
\hline
    \end{tabular}
    \caption{A comparative analysis of fundamental parameters for NGC 2345: our calculations with values reported in the literature.}
    \label{tab:fundamental parameters}
\end{table*}

\section{Dynamical Study} \label{sec:Dynamical Study}
\subsection{Luminosity and Mass function}

\begin{figure}[th]
    \centering
    \includegraphics[width= 9.5cm]{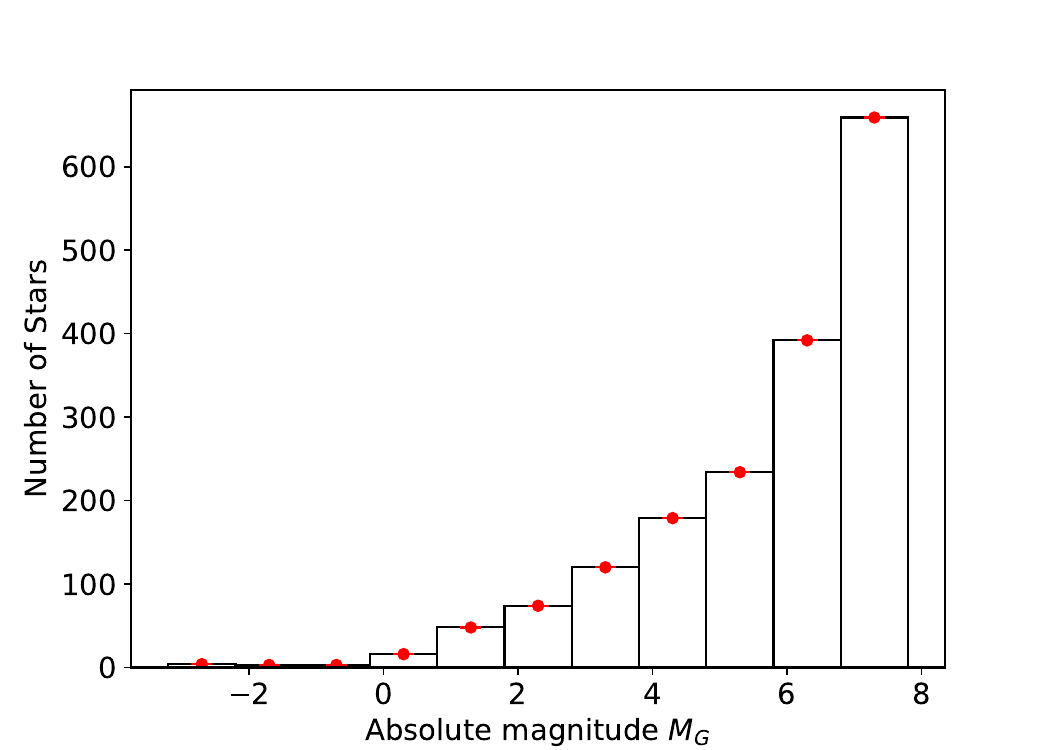}
    \caption{Histogram representing the distribution of main sequence stars within each specified magnitude bin, with $M_{G}$ denoting the absolute magnitude,  and red dots represent the associated errors.}
    \label{fig:Luminostiy_func_70prob.}
\end{figure}

The luminosity function characterizes the distribution of stars within a cluster according to their luminosity. The luminosity function for main-sequence stars within the cluster is derived from the calculated properties of the probable cluster stars. To construct the luminosity function, we only considered the stars having membership probability $\geq 70 \%$. The apparent G magnitude of these stars are converted to their absolute magnitude using the distance modulus and $A_{G}(=$1.66 mag). The magnitude bin interval of 1.0 mag was chosen to get a sufficient number of stars per magnitude bin for good statistics. Then, we constructed the histogram of LFs for the cluster, as shown in Fig \ref{fig:Luminostiy_func_70prob.}. We found an increasing luminosity function for this cluster. \\

\begin{figure}[th]
    \centering
    \includegraphics[width=9cm, height=8cm]{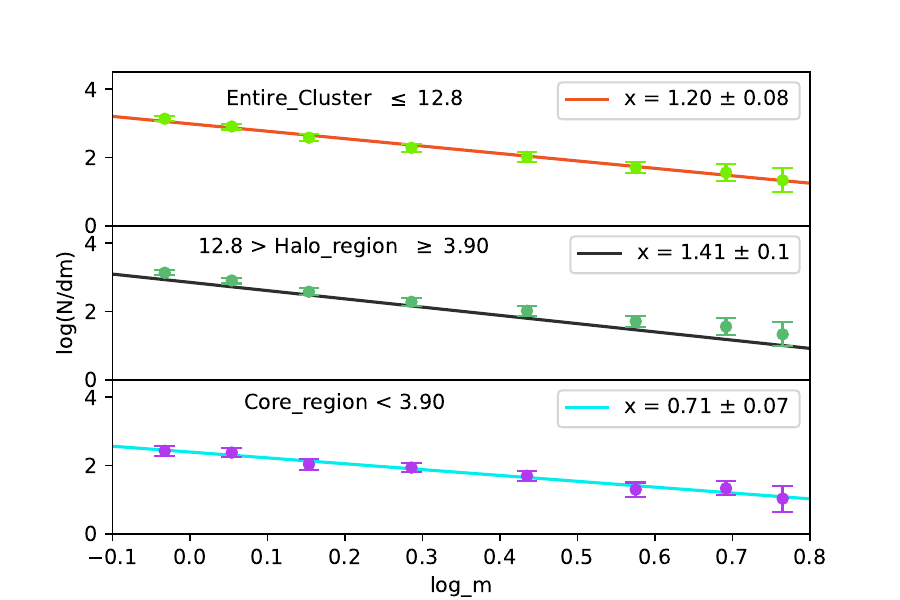}
    \caption{The mass function slope for NGC 2345 in three regions: core, halo, and cluster region. The slope of the mass function is calculated by least square fitting, as shown by solid lines in each panel.}
    \label{fig:70prob_mass_function_new}
\end{figure}

The mass function (MF) describes the distribution of masses among members of a cluster within a unit volume, and a mass-luminosity relationship can convert the LFs into the MF. Converting a cluster's luminosity functions (LFs) into the mass function involves utilizing best-fitted theoretical evolutionary tracks, as shown in Fig \ref{fig:70prob_mass_function_new}. 

We have utilized the theoretical models provided by \cite{marigo2017new} for this conversion. Through a least-square fitting approach, we determined the slope of the distribution by fitting the following equation:\\
\begin{equation}
    \log \frac{dN}{dM} = -(1+x) \times \log(M) + constant\\
\end{equation}
In the above relation, $dN$ is the number of stars in a particular mass bin $dM$, and $x$ is the slope of a mass function. The mass function slope was calculated for the cluster in three regions, i.e., core, halo, and entire cluster region. The parameters of the cluster are taken from the section \ref{sec:Structural Parameters of the Cluster}. The slope values of the mass function are listed in Table \ref{tab:mass function slope}. \\

\begin{table}

\caption{The mass function slope values for different cluster regions.}
\begin{center}
\begin{tabular}{|c|c|}
\hline
 Cluster &  NGC 2345 \\
\hline
 Mass Range $(M_\odot)$  & 0.84 $-$ 6.2       \\
\hline
 Mass Function Slope (x) &   \\
 Core &  0.70 $\pm$ 0.07  \\
 Halo &  1.41 $\pm$ 0.10 \\
 Entire Region & 1.20 $\pm$ 0.08 \\
 \hline
\end{tabular}
\end{center}
\label{tab:mass function slope}
\end{table}

The mass function slope is steeper for the core region, while for the entire cluster region, the slope is slightly lower than the Salpeter value of $x=1.354$ \citep{salpeter1955luminosity}. In contrast, the halo region exhibits good agreement with the Salpeter value. Furthermore, it is observed that the mass function slopes tend to increase as one moves toward the outer regions of the cluster compared to the core region. It may be because massive stars tend towards the cluster core while faint stars move towards the halo and outer region, and a possible hint of mass segregation is described in the next section.\\

\subsection{Mass Segregation}

\begin{figure}[th]
    \centering
    \includegraphics[width=9.5cm]{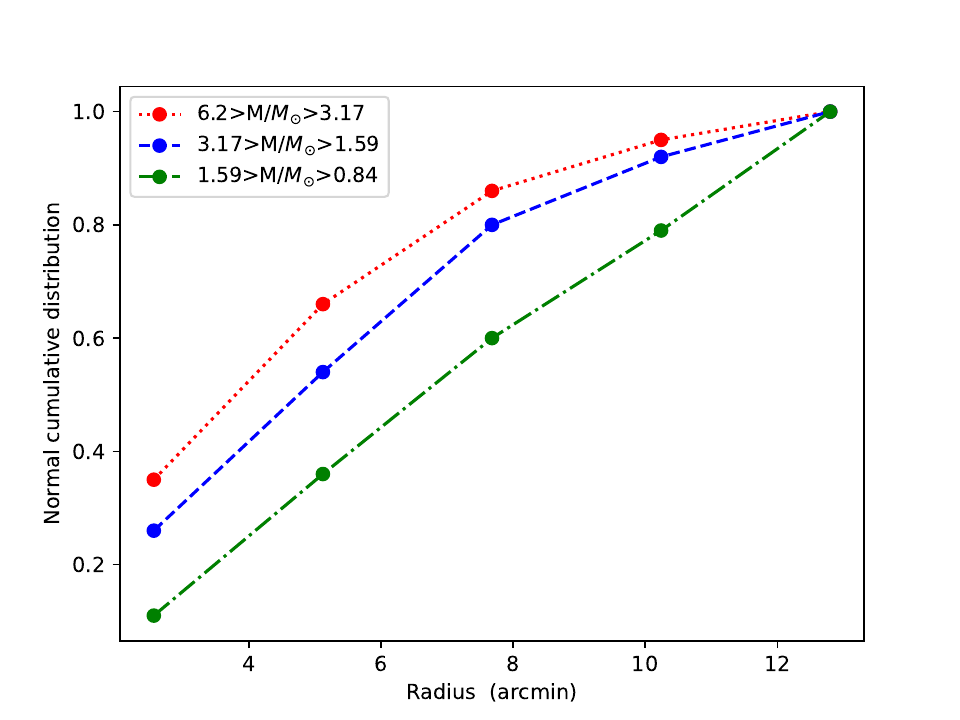}
    \caption{The cumulative radial distribution of stars in different mass ranges. The orange dashed line, blue dotted line, and green dashed dot line correspond to the high-, intermediate-, and low-mass ranges, respectively. The mass ranges are mentioned in the upper left of the graph.}
    \label{fig:cumulative_distribution_new}
\end{figure}

Open clusters are known to have mass segregation (\cite{dib2018structure}; \cite{dib2019star}). Mass segregation is the concentration of bright and massive stars towards the cluster's central region rather than the faint and low-mass stars. It is still a topic of debate whether mass segregation is primarily a result of the dynamical evolution of the cluster, the star formation process itself, or a combination of both (\cite{dib2018structure}; \cite{plunkett2018distribution}). During the dynamical evolution of clusters, massive stars undergo kinetic energy transfer to low-mass stars through the process of equipartition of energy, leading to an accumulation of massive stars in the central region of the cluster, while low-mass stars gradually migrate toward the outer region \citep{allison2009dynamical}. This process occurs because of the star formation process, which results in the preferential formation of massive stars in the central region of the cluster (\cite{dib2007origin}; \cite{dib2010imf}). Studying young open clusters is important for investigating the mass segregation process during dynamical evolution and its association with star formation \citep{pavlik2019star}. We create a cumulative distribution of stars based on their radial distance to study mass segregation within the cluster. We derived the cumulative radial stellar distribution of member stars for different mass ranges, as shown in Fig \ref{fig:cumulative_distribution_new}. We considered high-, intermediate-, and low-mass ranges, as indicated in Fig \ref{fig:cumulative_distribution_new}. We also conducted the Kolmogorov–Smirnov test on these mass ranges to determine whether they represent statistically different samples. We conclude with a confidence level of 80 percent that the mass segregation effect is present in the cluster NGC 2345.\\ 

\subsection{The Dynamical relaxation time}
The relaxation time provides a meaningful measure of the timescale over which a cluster will lose any remnants of its initial conditions. The relaxation timescale is characterized when a cluster reaches equipartition energy. The relaxation time given by \cite{spitzer1971random} is described by the following equation:\\
\begin{equation}
    T_{E} = \frac{8.9\times 10^{5}N^{1/2}R_{h}^{3/2}}{<M>^{1/2}\log(0.4N)} 
\end{equation}

Where N is the number of member stars, $R_{h}$ is the half mass radius (in pc) of the cluster, and   $<M>$ is the mean mass of the member stars in units of solar mass. The relaxation time in years, denoted as $T_{E}$, is determined based on the calculation of $R_{h}$, derived by considering the cumulative mass of stars as radial distance increases outward from the cluster center. $R_{h}$ represents the radial distance where half of the total cluster mass is contained.  We have calculated the value of $R_{h}$ based on the transformation equation as given in \citep{Larsen2006},\\

$R_{h}=0.547\times R_{c}\times \left (\frac{R_{t}}{R_{c}} \right)^{0.486}$\\

where $R_{c}$ is core radius while $R_{t}$ is tidal radius, and we have used those values from our estimation presented in this paper. We obtained $R_{h}$ value to be 8.65 $\pm$ 0.20 arcmin (6.97 $\pm$ 0.16pc) and the corresponding $T_{E}$ value as 177.6 Myr in the cluster. The cluster exhibits a relaxation time ($T_{E}$) that exceeds its current age; therefore, mass segregation in NGC 2345 prompts speculation regarding its potential association with the star formation process. Indeed, the early occurrence of mass segregation in clusters can be attributed to the rapid dynamical evolution that takes place, even within very young clusters, as suggested by certain previous studies (\cite{mcmillan2007dynamical}; \cite{allison2009dynamical}).  The cluster parameters found from the dynamical study of the open cluster NGC 2345 are listed in Table \ref{tab:dynamical parameters}.\\

\begin{table}[th]
\begin{center}
\caption{The calculated dynamical parameters for NGC 2345}
\begin{tabular}{|c|c|}
\hline
Cluster parameter    & NGC 2345 \\
\hline
Member stars     & 1732\\
Mean stellar mass $(M_\odot)$ & 1.826\\
Total mass $(M_\odot)$ & 3163\\
Cluster half radius $(R_{h}$/pc) &  6.97 $\pm 0.17$\\
Relaxation time $(T_{E}$/Myr)  &  177.6 \\
\hline
\end{tabular}
\label{tab:dynamical parameters}
\end{center}
\end{table}

\subsection{Apex of the Cluster}
The apex position shows the movement of star clusters across the celestial sphere. An Open Cluster (OC) is a gravitationally bound system of stars where the cluster members move with a shared velocity vector. Determining the apex position in OCs is paramount as it unveils crucial insights into the collective motion and dynamics of the cluster members across the celestial sphere. The methods described are based on the assumption that a cluster is a non-rotating body without any expansion or contraction \citep{maurya2021photometric}. The apex, obtained through the AD diagram method utilizing radial velocity and parallax measurements, is a critical parameter for understanding these clusters' kinematics, galactic dynamics, and formation history. It not only aids in deciphering the common origin of cluster members from a molecular cloud but also contributes to broader studies of the Milky Way's structure and the dynamics of stellar systems within it. We have used the AD diagram method to obtain the apex of the cluster. This method uses radial velocity and parallax of the stars. The AD diagram is discussed in detail by \cite{chupina2001geometry}, \cite{chupina2006kinematic}, \cite{vereshchagin2014apex}, \cite{elsanhoury2018pleiades}, and \cite{postnikova2020kinematical}. The (A, D) values of individual member stars indicate the positions of these stars through space velocity vectors. In this method, the intersection point $(A_{\o}, D_{\o})$, also referred to as the apex in equatorial coordinates, can be expressed as follows.
\begin{equation}
    A_{\o} = tan^{-1}\Biggl[\frac{\bar{V_{y}}}{\bar{V_{x}}}\Biggr], \hspace{0.2cm}   D_{\o} = tan^{-1}\Biggl[\frac{\bar{V_{z}}}{\sqrt{\bar{V_{x}^{2}}+\bar{V_{y}^{2}}}}\Biggr]
\end{equation}

\begin{figure}
    \centering
    \includegraphics[width=9.0cm,height=8cm]{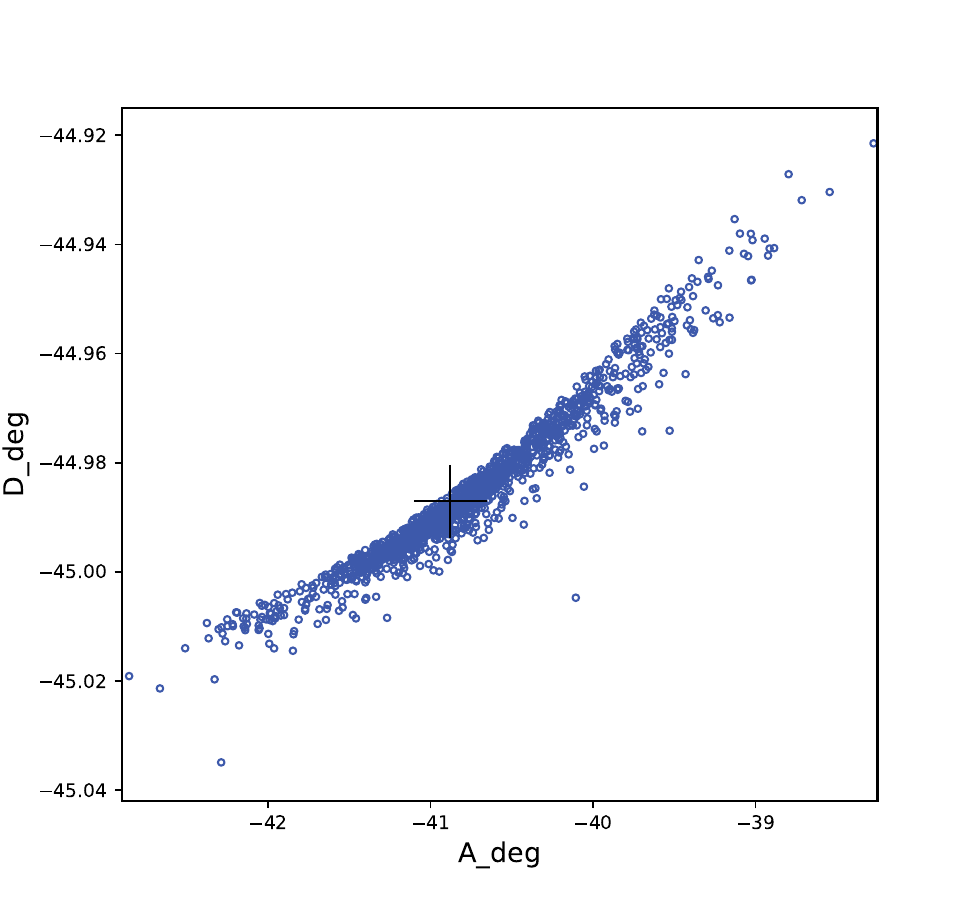}
    \caption{ AD-diagram for cluster NGC 2345. All stars (1732) are probable members with a membership probability higher than 70$\%$. A plus sign indicates the mean value of apex coordinates, described in the text.}
    \label{fig:apex_parameter}
\end{figure}

Where $V_{x}, V_{y}$ and $V_{z}$ represent the spatial velocities of stars on the celestial sphere. We have utilized the equation provided in \cite{chupina2001geometry} to calculate the velocity components. We determined the coordinates of the apex $(A_{\o}, D_{\o})$ using the AD diagram method, which were calculated as $(-40^{\circ}.89 \pm 0.12, -44^{\circ}.99 \pm 0.15)$.  Our estimated kinematical parameters are crucial for understanding the complete picture of a star's space motion within the cluster. This method uses the radial velocities and parallaxes of the \emph{Gaia} 1732 stars with membership
probability higher than 70$\%$ (Fig. \ref{fig:apex_parameter}).

\subsection{Motion of the clusters in the Galaxy}

To study the influence of the Galactic dynamic on the evolution of the cluster NGC 2345, we back-traced its motion in the Galaxy. 
Since the actual mass distribution of the Galaxy is unknown, we used the available mass models of the Galaxy. For this analysis, we used the most reliable Galactic potential model given by \cite{1991RMxAA..22..255A}. This model assumes that the total mass of the Galaxy is divided into three components: central bulge, disk, and halo. Potential for the bulge and disk are taken from \citet{1975PASJ...27..533M}, and for the halo region, potential from \citet{1999MNRAS.310..645W} is used. For this analysis, we used the updated parameters calculated by \cite{2017AstL...43..241B} using data from different sources up to a Galactocentric distance of 200 kpc. This model is widely used and discussed in many studies such as \cite{bisht2019mass} and \cite{2019MNRAS.490.1383R}. To perform the back integration of the orbital path
of the cluster in the Galaxy, we transformed the position and velocity components given in Table \ref{tab:fundamental parameters} 
of the cluster into Galactocentric coordinates. This analysis takes the cluster's radial velocity from \cite{alonso2019comprehensive} as 58.41 $\pm$ 0.15 km/s. We used the transformation matrices
given by  \cite{1987AJ.....93..864J}, coordinates of the Galactic north pole and Galactic center from \cite{2004ApJ...616..872R}. The result of this transformation, the Galactocentric position and velocity of the cluster are listed in Table \ref{tab:orbit_parameters}. The radial components are taken positively toward the Galactic Center, the tangential component is positive towards the Galactic rotation, and the vertical component is positive toward the Galactic North Pole.

\begin{figure*}[th]
    \centering
    \includegraphics[width=6.4cm, height=5.3cm]{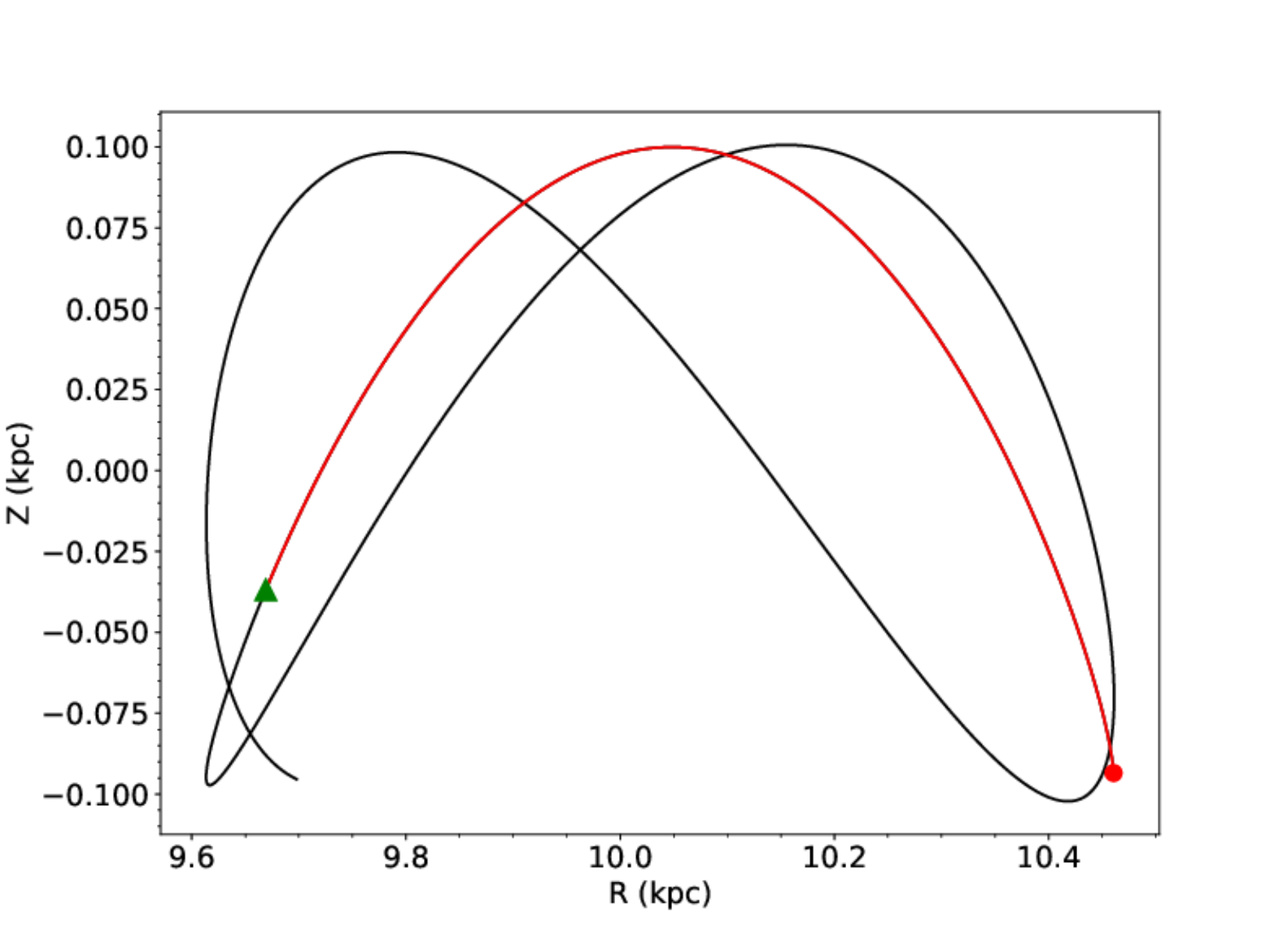}
    \includegraphics[width=5.3cm]{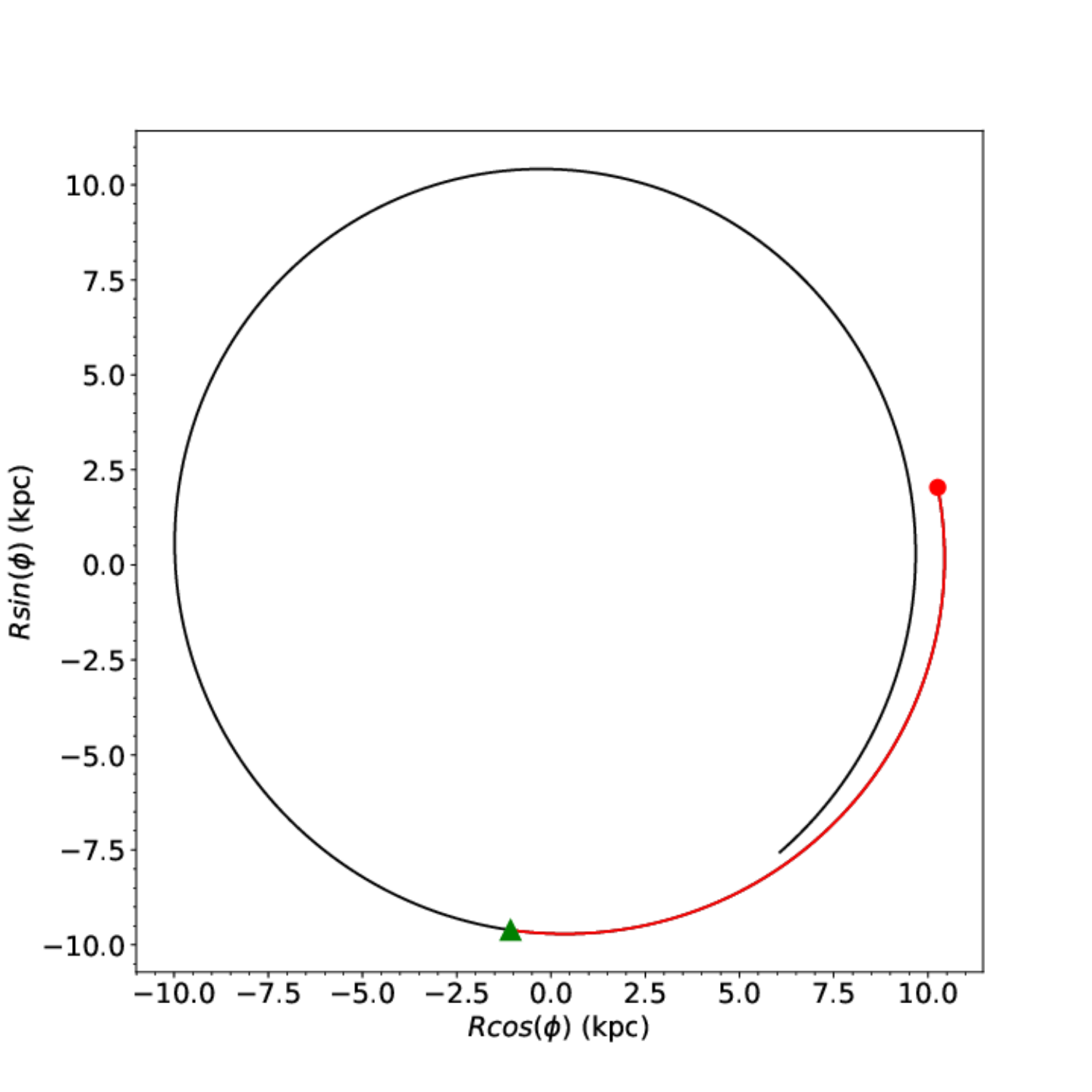}
    \includegraphics[width=6.4cm, height=5.3cm]{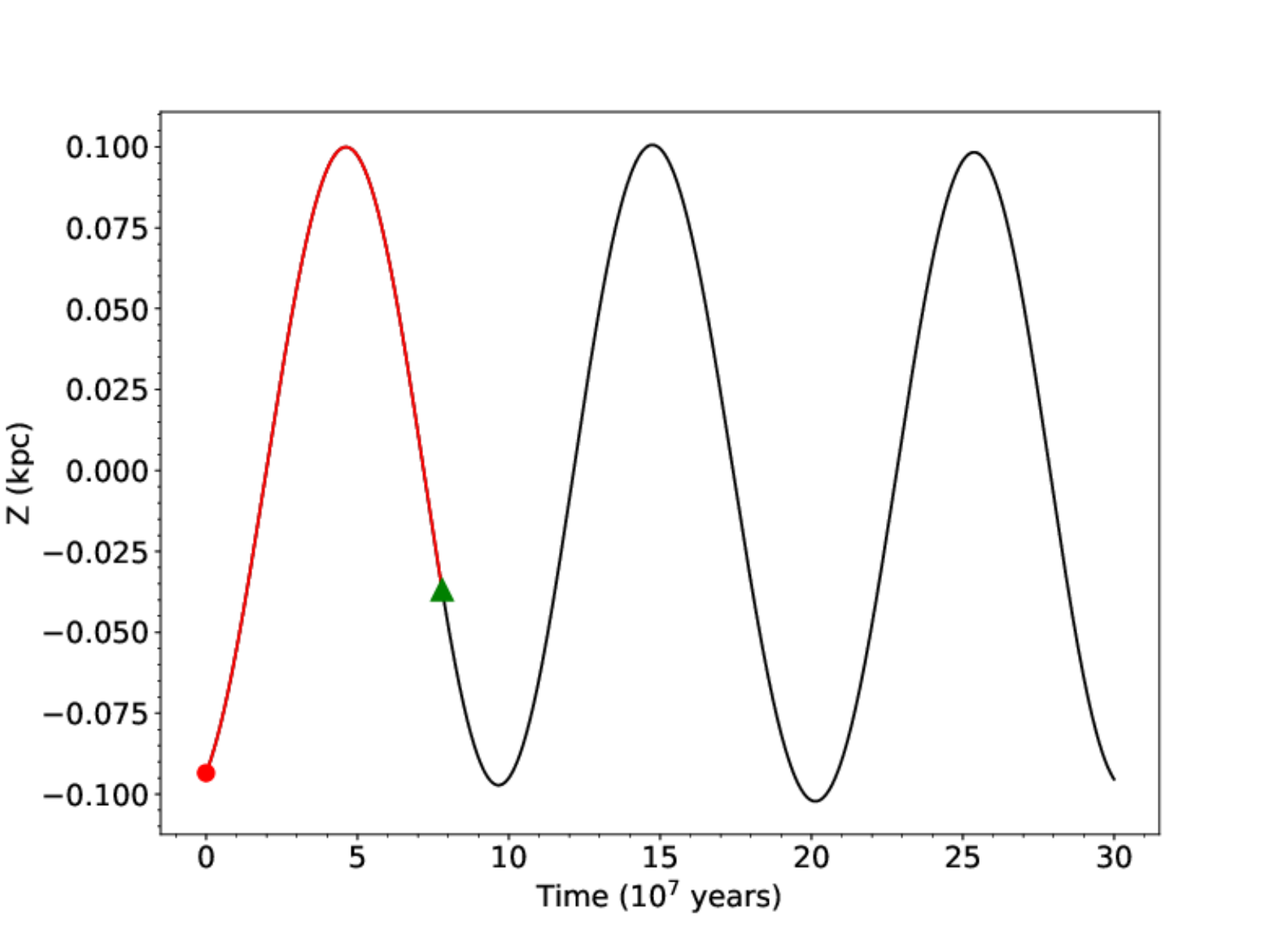}
    \caption{Galactic orbit of the NGC 2345 estimated with the Galactic potential model described in text in the time interval of the cluster's age. The top left panel shows a side view and the right panel shows a top view of the cluster's orbit. The bottom panel shows the motion of NGC 2345 in the Galactic disk with time. The filled triangle and filled circle denote the birth and present-day position of the cluster in the Galaxy.}
    \label{fig:orbits}
\end{figure*}

We back-integrated the orbits for a time equal to the cluster's age, as shown by the red curve in Fig. \ref{fig:orbits}. Due to a very young age, the cluster has not completed one revolution around the Galactic center; hence, we integrated the orbit for 300 Myr to determine the orbital properties of the cluster. The first panel from the left shows a plot between radial distance from the Galactic center and scale height from the Galactic disk. This plot shows that the cluster is orbiting outside the solar circle with a maximum scale height of 0.1 kpc, which makes it an object of thin Galactic disk. The green triangles in these plots represents the birth position, and the red circle is the present-day position of the cluster. It is visible that the cluster is born very close to the Galactic disk, hence highly affected by the Galactic tidal forces, which is visible as a close proximity of the cluster orbit to the Galactic disk. This close proximity will affect the evolution and survival of the cluster stars in the Galaxy. The increasing signature of the  luminosity function shown in Fig. \ref{fig:Luminostiy_func_70prob.} depicts that most of the cluster's low mass stars are bound to the cluster, which is due to its early age, but with time, we can predict that the stars will merge into the field distribution with a faster rate. The top right panel of Fig. \ref{fig:orbits} shows the plot between the components of the radial distance of the cluster. This plot shows that the cluster follows a circular path in the Galaxy while moving around the Galactic center. The bottom panel of this figure shows a plot between the time of motion and the perpendicular distance, which shows that the cluster is moving up to the same vertical height as the disk. This signifies that the cluster faces the same force across its path from the disk.

\begin{table}
\caption{This table presents the spacial and velocity coordinates of the cluster NGC 2345 in the Galactocentric coordinate system along with the orbital parameters of the cluster. R is the radial distance from the Galactic center, Z is the vertical distance of the cluster from the Galactic disk, and U, V, and W are the velocities of the clusters in radial, tangential, and vertical directions, respectively. $Z_{max}$ is the maximum distance the cluster travels from the Galactic disk. R$_{A}$ and R$_{P}$ apogalactic and perigalactic radii of the orbit, respectively. L$_{Z}$ is the third component of the angular momentum.}
\begin{center}
\begin{tabular}{|l|c|}
\hline
Parameter & NGC 2345 \\
\hline
(R,Z) (kpc) & (10.46 $\pm$ 0.79, -0.09 $\pm$ 0.04) \\
U (km/s) & -0.53$\pm$7.22   \\
V (km/s)  & -235.16 $\pm$ 7.15 \\
W (km/s)  & 2.41 $\pm$ 7.14 \\
eccentricity & 0.001 \\
(R$_{A}$, R$_{P}$) (kpc) & (10.44, 10.42) \\
Z$_{max}$ (kpc) &  0.10 \\
Energy (100 km/s)$^{2}$ & -09.85  \\
L$_{Z}$ (100 kpc km/s) &   -24.60 \\
Time period (Myr)  & 278  \\
\hline
\end{tabular}
\end{center}
\label{tab:orbit_parameters}
\end{table}

\section{Summary and Conclusions}
\label{sec:Summary and conclusion}

We thoroughly analyze the young open cluster NGC 2345 through photometric and kinematic studies. Our investigation is based on \emph{UBV} data from the 2.0m Himalayan Chandra telescope, complemented by valuable insights from \emph{Gaia} DR3, 2MASS, and APASS Data. The findings of our study are summarized as follows:
\begin{enumerate}
    \item We have estimated the cluster center with high precision using the cluster members, found the following coordinates: $\alpha$ = 107.08 $\pm$ 0.07 degrees ($07^{h} 08^{m} 8.3^{s}$), and $\delta$ = -13.20 $\pm$ 0.08  degrees ($-13^{\circ} 11^{\prime} 38^{\prime\prime} $).
    \item By analyzing the radial density profile, we determined the cluster radius to be 12.8 arcminutes, equivalent to approximately 10.37 parsecs (adopting a distance of 2.78 kpc). We found the cluster core and tidal radius as $\sim$ 3.9 and 20 arcmin.
    \item Using \emph{Gaia} data, we have estimated the membership probability and find 1732 most probable cluster members for NGC 2345 with membership probability higher than 70 $\%$. The estimated mean proper motion as $-1.34 \pm 0.20$ mas $yr^{-1}$ and $1.35 \pm 0.21$ mas $yr^{-1}$ in both the directions of RA and DEC, respectively.
    \item From the two color diagram, we have estimated $E(B-V) = 0.63 \pm 0.04$ mag. The combination of 2MASS $JHK$ data and optical data provides $E(J-H) = 0.20\pm 0.04$ mag and $E(J-K) = 0.34\pm 0.03$ mag for NGC 2345. We found that the interstellar extinction law is normal towards the cluster region.
    \item The distance to NGC 2345 was determined using the Bailer-Jones method and found to be 2.78 $\pm$ 0.780 kpc. Additionally, we estimated the distance using the cluster's true distance modulus, resulting in a value of 2.51 $\pm$ 0.12 kpc. The age has been determined to be $63 \pm 8 $ million years (Myr) through a comparative analysis of the cluster's Color-Magnitude Diagram (CMD) with metallicity $Z = 0.01$ theoretical isochrones provided by \cite{marigo2017new}.
    \item We have identified the Mass Function (MF) slope to be $1.2 \pm 0.08$ for all stars located within the complete cluster radius. We found a signature of mass segregation based on slopes in the cluster's core, halo, and overall regions.
    \item We found that the dynamical relaxation time for NGC 2345 is larger than the cluster's age. Thus, the dynamical evolution process is ongoing in the cluster; after 100 Myr, it will be dynamically relaxed.
   
    \item We derived the kinematic parameters based on the radial velocity and parallax of the cluster. Utilizing the AD diagram, we determined the apex position to be $(-40^{\circ}.89 \pm 0.12, -44^{\circ}.99 \pm 0.15)$.
    
    \item We studied the effect of Galaxy dynamics on the cluster NGC 2345 by studying its motion in the Galaxy.
    This shows that the cluster is moving in a circular path and experiencing the same amount of force from the Galactic disk across its journey around the Galactic center. The cluster has not yet completed a single orbit around the Galactic center.
    
\end{enumerate}

\section*{Acknowledgements}
 We sincerely thank the anonymous reviewer's generous time and expertise, greatly improving the manuscript's quality. We are thankful to the observers of the 2-m HCT for their contributions to accumulating photometric data of this cluster. This work has made use of data from the European Space Agency (ESA) mission \emph{Gaia} (https://www.cosmos.esa.int/gaia), processed by the \emph{Gaia} Data Processing and Analysis Consortium (DPAC, https: //www.cosmos.esa.int/web/gaia/dpac/consortium). The DPAC (Data Processing and Analysis Consortium) funding has been provided by national institutions, focusing on those participating in the \emph{Gaia} Multilateral Agreement. Additionally, this work has utilized WEBDA and data products from the Two Micron All Sky Survey (2MASS), a collaborative project between the University of Massachusetts and the Infrared Processing and Analysis Center/California Institute of Technology. The 2MASS project is funded by the National Aeronautics and Space Administration (NASA) and the National Science Foundation (NSF).

\bibliography{main}{}

\begin{thebibliography}{}
\expandafter\ifx\csname natexlab\endcsname\relax\def\natexlab#1{#1}\fi
\providecommand{\url}[1]{\href{#1}{#1}}
\providecommand{\dodoi}[1]{doi:~\href{http://doi.org/#1}{\nolinkurl{#1}}}
\providecommand{\doeprint}[1]{\href{http://ascl.net/#1}{\nolinkurl{http://ascl.net/#1}}}
\providecommand{\doarXiv}[1]{\href{https://arxiv.org/abs/#1}{\nolinkurl{https://arxiv.org/abs/#1}}}

\bibitem[{{Allen} \& {Santillan}(1991)}]{1991RMxAA..22..255A}
{Allen}, C., \& {Santillan}, A. 1991, \rmxaa, 22, 255

\bibitem[{Allison {et~al.}(2009)Allison, Goodwin, Parker, De~Grijs, Zwart, \& Kouwenhoven}]{allison2009dynamical}
Allison, R.~J., Goodwin, S.~P., Parker, R.~J., {et~al.} 2009, The Astrophysical Journal, 700, L99

\bibitem[{Alonso-Santiago {et~al.}(2019)Alonso-Santiago, Negueruela, Marco, Tabernero, Gonz{\'a}lez-Fern{\'a}ndez, \& Castro}]{alonso2019comprehensive}
Alonso-Santiago, J., Negueruela, I., Marco, A., {et~al.} 2019, Astronomy \& Astrophysics, 631, A124

\bibitem[{{Astraatmadja} \& {Bailer-Jones}(2016)}]{2016ApJ...833..119A}
{Astraatmadja}, T.~L., \& {Bailer-Jones}, C. A.~L. 2016, \apj, 833, 119, \dodoi{10.3847/1538-4357/833/1/119}

\bibitem[{Bailer-Jones {et~al.}(2018)Bailer-Jones, Rybizki, Fouesneau, Mantelet, \& Andrae}]{bailer2018estimating}
Bailer-Jones, C., Rybizki, J., Fouesneau, M., Mantelet, G., \& Andrae, R. 2018, The Astronomical Journal, 156, 58

\bibitem[{Bailer-Jones(2015)}]{bailer2015estimating}
Bailer-Jones, C.~A. 2015, Publications of the Astronomical Society of the Pacific, 127, 994

\bibitem[{Balaguer-N{\'u}nez {et~al.}(1998)Balaguer-N{\'u}nez, Tian, \& Zhao}]{balaguer1998determination}
Balaguer-N{\'u}nez, L., Tian, K., \& Zhao, J. 1998, Astronomy and Astrophysics Supplement Series, 133, 387

\bibitem[{Bastian {et~al.}(2010)Bastian, Covey, \& Meyer}]{bastian2010universal}
Bastian, N., Covey, K.~R., \& Meyer, M.~R. 2010, Annual Review of Astronomy and Astrophysics, 48, 339

\bibitem[{Becker \& Stock(1954)}]{becker1954drei}
Becker, W., \& Stock, J. 1954, Zeitschrift f{\"u}r Astrophysik, Vol. 34, p. 1, 34, 1

\bibitem[{Bisht {et~al.}(2019)Bisht, Yadav, Ganesh, Durgapal, Rangwal, \& Fynbo}]{bisht2019mass}
Bisht, D., Yadav, R., Ganesh, S., {et~al.} 2019, Monthly Notices of the Royal Astronomical Society, 482, 1471

\bibitem[{Bisht {et~al.}(2022{\natexlab{a}})Bisht, Zhu, Elsanhoury, Yadav, Rangwal, Sariya, Durgapal, \& Jiang}]{bisht2022comprehensive}
Bisht, D., Zhu, Q., Elsanhoury, W., {et~al.} 2022{\natexlab{a}}, The Astronomical Journal, 164, 171

\bibitem[{Bisht {et~al.}(2021{\natexlab{a}})Bisht, Zhu, Elsanhoury, Sariya, Rangwal, Yadav, Durgapal, \& Jiang}]{bisht2021detailed}
Bisht, D., Zhu, Q., Elsanhoury, W.~H., {et~al.} 2021{\natexlab{a}}, Publications of the Astronomical Society of Japan, 73, 677

\bibitem[{Bisht {et~al.}(2020)Bisht, Zhu, Yadav, Durgapal, \& Rangwal}]{bisht2020comprehensive}
Bisht, D., Zhu, Q., Yadav, R., Durgapal, A., \& Rangwal, G. 2020, Monthly Notices of the Royal Astronomical Society, 494, 607

\bibitem[{Bisht {et~al.}(2021{\natexlab{b}})Bisht, Zhu, Yadav, Rangwal, Durgapal, Sariya, \& Jiang}]{bisht2021deep}
Bisht, D., Zhu, Q., Yadav, R., {et~al.} 2021{\natexlab{b}}, The Astronomical Journal, 161, 182

\bibitem[{Bisht {et~al.}(2022{\natexlab{b}})Bisht, Zhu, Yadav, Rangwal, Sariya, Durgapal, \& Jiang}]{bisht2022deep11}
---. 2022{\natexlab{b}}, Publications of the Astronomical Society of the Pacific, 134, 044201

\bibitem[{{Bobylev} {et~al.}(2017){Bobylev}, {Bajkova}, \& {Gromov}}]{2017AstL...43..241B}
{Bobylev}, V.~V., {Bajkova}, A.~T., \& {Gromov}, A.~O. 2017, Astronomy Letters, 43, 241, \dodoi{10.1134/S1063773717040016}

\bibitem[{Caldwell {et~al.}(1993)Caldwell, Cousins, Ahlers, Van~Wamelen, \& Maritz}]{caldwell1993statistical}
Caldwell, J.~A., Cousins, A., Ahlers, C., Van~Wamelen, P., \& Maritz, E. 1993, SAAO Circulars, Vol. 15, p. 1, 15, 1

\bibitem[{Cantat-Gaudin {et~al.}(2018)Cantat-Gaudin, Jordi, Vallenari, Bragaglia, Balaguer-N{\'u}{\~n}ez, Soubiran, Bossini, Moitinho, Castro-Ginard, Krone-Martins, {et~al.}}]{cantat2018gaia}
Cantat-Gaudin, T., Jordi, C., Vallenari, A., {et~al.} 2018, Astronomy \& Astrophysics, 618, A93

\bibitem[{Cantat-Gaudin {et~al.}(2019)Cantat-Gaudin, Krone-Martins, Sedaghat, Farahi, de~Souza, Skalidis, Malz, Mac{\^e}do, Moews, Jordi, {et~al.}}]{cantat2019gaia}
Cantat-Gaudin, T., Krone-Martins, A., Sedaghat, N., {et~al.} 2019, Astronomy \& Astrophysics, 624, A126

\bibitem[{Cantat-Gaudin {et~al.}(2020)Cantat-Gaudin, Anders, Castro-Ginard, Jordi, Romero-G{\'o}mez, Soubiran, Casamiquela, Tarricq, Moitinho, Vallenari, {et~al.}}]{cantat2020painting}
Cantat-Gaudin, T., Anders, F., Castro-Ginard, A., {et~al.} 2020, Astronomy \& Astrophysics, 640, A1

\bibitem[{Cardelli {et~al.}(1989)Cardelli, Clayton, \& Mathis}]{cardelli1989relationship}
Cardelli, J.~A., Clayton, G.~C., \& Mathis, J.~S. 1989, Astrophysical Journal, Part 1 (ISSN 0004-637X), vol. 345, Oct. 1, 1989, p. 245-256., 345, 245

\bibitem[{Carraro {et~al.}(2014)Carraro, V{\'a}zquez, Costa, Ahumada, \& Giorgi}]{carraro2014thickening}
Carraro, G., V{\'a}zquez, R.~A., Costa, E., Ahumada, J.~A., \& Giorgi, E.~E. 2014, The Astronomical Journal, 149, 12

\bibitem[{Carraro {et~al.}(2008)Carraro, Villanova, Demarque, Bidin, \& McSwain}]{carraro2008old}
Carraro, G., Villanova, S., Demarque, P., Bidin, C.~M., \& McSwain, M. 2008, Monthly Notices of the Royal Astronomical Society, 386, 1625

\bibitem[{Carrera {et~al.}(2022)Carrera, Casamiquela, Bragaglia, Carretta, Carbajo-Hijarrubia, Jordi, Alonso-Santiago, Balaguer-Nu{\~n}ez, Baratella, D’Orazi, {et~al.}}]{carrera2022one}
Carrera, R., Casamiquela, L., Bragaglia, A., {et~al.} 2022, Astronomy \& Astrophysics, 663, A148

\bibitem[{Castro-Ginard {et~al.}(2019)Castro-Ginard, Jordi, Luri, Cantat-Gaudin, \& Balaguer-N{\'u}{\~n}ez}]{castro2019hunting}
Castro-Ginard, A., Jordi, C., Luri, X., Cantat-Gaudin, T., \& Balaguer-N{\'u}{\~n}ez, L. 2019, Astronomy \& Astrophysics, 627, A35

\bibitem[{Castro-Ginard {et~al.}(2018)Castro-Ginard, Jordi, Luri, Julbe, Morvan, Balaguer-N{\'u}{\~n}ez, \& Cantat-Gaudin}]{castro2018new}
Castro-Ginard, A., Jordi, C., Luri, X., {et~al.} 2018, Astronomy \& Astrophysics, 618, A59

\bibitem[{Chini {et~al.}(1990)Chini, Kr{\"u}gel, \& Kreysa}]{chini1990large}
Chini, R., Kr{\"u}gel, E., \& Kreysa, E. 1990, Astronomy and Astrophysics (ISSN 0004-6361), vol. 227, no. 1, Jan. 1990, p. L5-L8., 227, L5

\bibitem[{Chupina {et~al.}(2001)Chupina, Reva, \& Vereshchagin}]{chupina2001geometry}
Chupina, N., Reva, V., \& Vereshchagin, S. 2001, Astronomy \& Astrophysics, 371, 115

\bibitem[{Chupina {et~al.}(2006)Chupina, Reva, \& Vereshchagin}]{chupina2006kinematic}
---. 2006, Astronomy \& Astrophysics, 451, 909

\bibitem[{Dias {et~al.}(2018)Dias, Monteiro, L{\'e}pine, Prates, Gneiding, \& Sacchi}]{dias2018astrometric}
Dias, W.~S., Monteiro, H., L{\'e}pine, J. R.~D., {et~al.} 2018, Monthly Notices of the Royal Astronomical Society, 481, 3887

\bibitem[{Dias {et~al.}(2021)Dias, Monteiro, Moitinho, L{\'e}pine, Carraro, Paunzen, Alessi, \& Villela}]{dias2021updated}
Dias, W.~S., Monteiro, H., Moitinho, A., {et~al.} 2021, Monthly Notices of the Royal Astronomical Society, 504, 356

\bibitem[{Dib \& Basu(2018)}]{dib2018emergence}
Dib, S., \& Basu, S. 2018, Astronomy \& Astrophysics, 614, A43

\bibitem[{Dib \& Henning(2019)}]{dib2019star}
Dib, S., \& Henning, T. 2019, Astronomy \& Astrophysics, 629, A135

\bibitem[{Dib {et~al.}(2007)Dib, Kim, \& Shadmehri}]{dib2007origin}
Dib, S., Kim, J., \& Shadmehri, M. 2007, Monthly Notices of the Royal Astronomical Society: Letters, 381, L40

\bibitem[{Dib {et~al.}(2018)Dib, Schmeja, \& Parker}]{dib2018structure}
Dib, S., Schmeja, S., \& Parker, R.~J. 2018, Monthly Notices of the Royal Astronomical Society, 473, 849

\bibitem[{Dib {et~al.}(2010)Dib, Shadmehri, Padoan, Maheswar, Ojha, \& Khajenabi}]{dib2010imf}
Dib, S., Shadmehri, M., Padoan, P., {et~al.} 2010, Monthly Notices of the Royal Astronomical Society, 405, 401

\bibitem[{Elsanhoury {et~al.}(2018)Elsanhoury, Postnikova, Chupina, Vereshchagin, Sariya, Yadav, \& Jiang}]{elsanhoury2018pleiades}
Elsanhoury, W., Postnikova, E., Chupina, N., {et~al.} 2018, Astrophysics and space science, 363, 1

\bibitem[{Fiorucci \& Munari(2003)}]{fiorucci2003asiago}
Fiorucci, M., \& Munari, U. 2003, Astronomy \& Astrophysics, 401, 781

\bibitem[{Girard {et~al.}(1989)Girard, Grundy, L{\'o}pez, \& van Altena}]{girard1989relative}
Girard, T.~M., Grundy, W.~M., L{\'o}pez, C.~E., \& van Altena, W.~F. 1989, Astronomical Journal (ISSN 0004-6256), vol. 98, July 1989, p. 227-243. Research supported by NSF., 98, 227

\bibitem[{Groenewegen(2021)}]{groenewegen2021parallax}
Groenewegen, M. 2021, Astronomy \& Astrophysics, 654, A20

\bibitem[{Henden \& Munari(2014)}]{henden2014apass}
Henden, A., \& Munari, U. 2014, Contrib. Astron. Obs. Skalnate Pleso, 43, 518

\bibitem[{{Johnson} \& {Soderblom}(1987)}]{1987AJ.....93..864J}
{Johnson}, D.~R.~H., \& {Soderblom}, D.~R. 1987, \aj, 93, 864, \dodoi{10.1086/114370}

\bibitem[{Kalirai \& Tosi(2004)}]{kalirai2004interpreting}
Kalirai, J.~S., \& Tosi, M. 2004, Monthly Notices of the Royal Astronomical Society, 351, 649

\bibitem[{Kharchenko {et~al.}(2005)Kharchenko, Piskunov, R{\"o}ser, Schilbach, \& Scholz}]{kharchenko2005astrophysical}
Kharchenko, N., Piskunov, A., R{\"o}ser, S., Schilbach, E., \& Scholz, R.-D. 2005, Astronomy \& Astrophysics, 438, 1163

\bibitem[{Kharchenko {et~al.}(2013)Kharchenko, Piskunov, Schilbach, R{\"o}ser, \& Scholz}]{kharchenko2013global}
Kharchenko, N., Piskunov, A., Schilbach, E., R{\"o}ser, S., \& Scholz, R.-D. 2013, Astronomy \& Astrophysics, 558, A53

\bibitem[{King(1962)}]{king1962structure}
King, I. 1962, Astronomical Journal, Vol. 67, p. 471 (1962), 67, 471

\bibitem[{Landolt(1992)}]{landolt1992ubvri}
Landolt, A.~U. 1992, Astronomical Journal (ISSN 0004-6256), vol. 104, no. 1, July 1992, p. 340-371, 436-491. Research supported by Space Telescope Science Institute., 104, 340

\bibitem[{Larsen(2006)}]{Larsen2006}
Larsen, S.~S. 2006, An ISHAPE Users Guide 14, arXiv:astro-ph/0701774

\bibitem[{Liu \& Pang(2019)}]{liu2019catalog}
Liu, L., \& Pang, X. 2019, The Astrophysical Journal Supplement Series, 245, 32

\bibitem[{Luri {et~al.}(2018)Luri, Brown, Sarro, Arenou, Bailer-Jones, Castro-Ginard, de~Bruijne, Prusti, Babusiaux, \& Delgado}]{luri2018gaia}
Luri, X., Brown, A., Sarro, L., {et~al.} 2018, Astronomy \& Astrophysics, 616, A9

\bibitem[{Marigo {et~al.}(2017)Marigo, Girardi, Bressan, Rosenfield, Aringer, Chen, Dussin, Nanni, Pastorelli, Rodrigues, {et~al.}}]{marigo2017new}
Marigo, P., Girardi, L., Bressan, A., {et~al.} 2017, The Astrophysical Journal, 835, 77

\bibitem[{Maurya {et~al.}(2021)Maurya, Joshi, Elsanhoury, \& Sharma}]{maurya2021photometric}
Maurya, J., Joshi, Y., Elsanhoury, W., \& Sharma, S. 2021, The Astronomical Journal, 162, 64

\bibitem[{Maurya {et~al.}(2020)Maurya, Joshi, \& Gour}]{maurya2020photometric}
Maurya, J., Joshi, Y., \& Gour, A. 2020, Monthly Notices of the Royal Astronomical Society, 495, 2496

\bibitem[{McMillan {et~al.}(2007)McMillan, Vesperini, \& Zwart}]{mcmillan2007dynamical}
McMillan, S.~L., Vesperini, E., \& Zwart, S. F.~P. 2007, The Astrophysical Journal, 655, L45

\bibitem[{{Miyamoto} \& {Nagai}(1975)}]{1975PASJ...27..533M}
{Miyamoto}, M., \& {Nagai}, R. 1975, \pasj, 27, 533

\bibitem[{Moffat(1974)}]{moffat1974ngc}
Moffat, A. 1974, Astronomy and Astrophysics Suppl. Vol. 16, p. 33, 16, 33

\bibitem[{Monteiro \& Dias(2019)}]{monteiro2019distances}
Monteiro, H., \& Dias, W. 2019, Monthly Notices of the Royal Astronomical Society, 487, 2385

\bibitem[{Pavl{\'\i}k {et~al.}(2019)Pavl{\'\i}k, Kroupa, \& {\v{S}}ubr}]{pavlik2019star}
Pavl{\'\i}k, V., Kroupa, P., \& {\v{S}}ubr, L. 2019, Astronomy \& Astrophysics, 626, A79

\bibitem[{Plunkett {et~al.}(2018)Plunkett, Fern{\'a}ndez-L{\'o}pez, Arce, Busquet, Mardones, \& Dunham}]{plunkett2018distribution}
Plunkett, A.~L., Fern{\'a}ndez-L{\'o}pez, M., Arce, H.~G., {et~al.} 2018, Astronomy \& Astrophysics, 615, A9

\bibitem[{Portegies~Zwart {et~al.}(2010)Portegies~Zwart, McMillan, \& Gieles}]{portegies2010young}
Portegies~Zwart, S.~F., McMillan, S.~L., \& Gieles, M. 2010, Annual review of astronomy and astrophysics, 48, 431

\bibitem[{Postnikova {et~al.}(2020)Postnikova, Elsanhoury, Sariya, Chupina, Vereshchagin, \& Jiang}]{postnikova2020kinematical}
Postnikova, E., Elsanhoury, W., Sariya, D.~P., {et~al.} 2020, Research in Astronomy and Astrophysics, 20, 016

\bibitem[{{Rangwal} {et~al.}(2019){Rangwal}, {Yadav}, {Durgapal}, {Bisht}, \& {Nardiello}}]{2019MNRAS.490.1383R}
{Rangwal}, G., {Yadav}, R.~K.~S., {Durgapal}, A., {Bisht}, D., \& {Nardiello}, D. 2019, \mnras, 490, 1383, \dodoi{10.1093/mnras/stz2642}

\bibitem[{{Reid} \& {Brunthaler}(2004)}]{2004ApJ...616..872R}
{Reid}, M.~J., \& {Brunthaler}, A. 2004, \apj, 616, 872, \dodoi{10.1086/424960}

\bibitem[{Salpeter(1955)}]{salpeter1955luminosity}
Salpeter, E.~E. 1955, Astrophysical Journal, vol. 121, p. 161, 121, 161

\bibitem[{Sariya {et~al.}(2021)Sariya, Jiang, Sizova, Postnikova, Bisht, Chupina, Vereshchagin, Yadav, Rangwal, \& Tutukov}]{sariya2021comprehensive}
Sariya, D.~P., Jiang, G., Sizova, M., {et~al.} 2021, The Astronomical Journal, 161, 101

\bibitem[{Schmid-Kaler(1982)}]{schmid1982landolt}
Schmid-Kaler, T. 1982, in New series, Group VI, Vol.~2 (Springer-Verlag Berlin), 14

\bibitem[{Sharma {et~al.}(2020)Sharma, Ghosh, Ojha, Pandey, Sinha, Pandey, Ghosh, Panwar, \& Pandey}]{sharma2020disintegrating}
Sharma, S., Ghosh, A., Ojha, D., {et~al.} 2020, Monthly Notices of the Royal Astronomical Society, 498, 2309

\bibitem[{Singh {et~al.}(2022)Singh, Pandey, \& Hoang}]{singh2022polarization}
Singh, S., Pandey, J.~C., \& Hoang, T. 2022, Monthly Notices of the Royal Astronomical Society, 513, 4899

\bibitem[{Sneden {et~al.}(1978)Sneden, Gehrz, Hackwell, York, \& Snow}]{sneden1978infrared}
Sneden, C., Gehrz, R., Hackwell, J., York, D., \& Snow, T. 1978, The Astrophysical Journal, 223, 168

\bibitem[{Spitzer~Jr \& Hart(1971)}]{spitzer1971random}
Spitzer~Jr, L., \& Hart, M.~H. 1971, Astrophysical Journal, vol. 166, p. 483, 166, 483

\bibitem[{Stalin {et~al.}(2008)Stalin, Hegde, Sahu, Parihar, Anupama, Bhatt, \& Prabhu}]{stalin2008night}
Stalin, C., Hegde, M., Sahu, D., {et~al.} 2008, arXiv preprint arXiv:0809.1745

\bibitem[{Stetson(1987)}]{stetson1987daophot}
Stetson, P.~B. 1987, Publications of the Astronomical Society of the Pacific, 99, 191

\bibitem[{Sun {et~al.}(2021)Sun, de~Grijs, Deng, \& Albrow}]{sun2021binary}
Sun, W., de~Grijs, R., Deng, L., \& Albrow, M.~D. 2021, Monthly Notices of the Royal Astronomical Society, 502, 4350

\bibitem[{Tapia {et~al.}(1988)Tapia, Roth, Marraco, \& Ruiz}]{tapia1988interstellar}
Tapia, M., Roth, M., Marraco, H., \& Ruiz, M. 1988, Monthly Notices of the Royal Astronomical Society, 232, 661

\bibitem[{Tsantaki {et~al.}(2023)Tsantaki, Delgado-Mena, Bossini, Sousa, Pancino, \& Martins}]{tsantaki2023search}
Tsantaki, M., Delgado-Mena, E., Bossini, D., {et~al.} 2023, Astronomy \& Astrophysics, 674, A157

\bibitem[{Vallenari {et~al.}(2023)Vallenari, Brown, Prusti, de~Bruijne, Arenou, Babusiaux, Biermann, Creevey, Ducourant, Evans, {et~al.}}]{vallenari2023gaia}
Vallenari, A., Brown, A., Prusti, T., {et~al.} 2023, Astronomy \& Astrophysics, 674, A1

\bibitem[{Vereshchagin {et~al.}(2014)Vereshchagin, Chupina, Sariya, Yadav, \& Kumar}]{vereshchagin2014apex}
Vereshchagin, S., Chupina, N., Sariya, D.~P., Yadav, R., \& Kumar, B. 2014, New Astronomy, 31, 43

\bibitem[{{Wilkinson} \& {Evans}(1999)}]{1999MNRAS.310..645W}
{Wilkinson}, M.~I., \& {Evans}, N.~W. 1999, \mnras, 310, 645, \dodoi{10.1046/j.1365-8711.1999.02964.x}

\bibitem[{Worrall {et~al.}(1992)Worrall, Biemesderfer, \& Barnes}]{worrall1992astronomical}
Worrall, D.~M., Biemesderfer, C., \& Barnes, J. 1992, Astronomical Data Analysis Software and Systems I, 25

\bibitem[{Yadav {et~al.}(2013)Yadav, Sariya, \& Sagar}]{yadav2013proper}
Yadav, R., Sariya, D.~P., \& Sagar, R. 2013, Monthly Notices of the Royal Astronomical Society, 430, 3350

\bibitem[{Yontan(2023)}]{yontan2023investigation}
Yontan, T. 2023, The Astronomical Journal, 165, 79

\end{thebibliography}
\bibliographystyle{aasjournal}

\end{document}